\documentclass[11pt,a4paper,twoside,openright]{report}

\usepackage[utf8]{inputenc}

\usepackage{float}

\usepackage[mieic,final]{feupteses}

\usepackage[lofdepth,lotdepth]{subfig}
\usepackage{graphicx}

\usepackage{acronym}
\usepackage{amsmath}
\usepackage{amssymb}
\usepackage{pifont}
\usepackage{gantt}
\usepackage{wasysym}
\usepackage{array}
\usepackage{pdfpages}

\usepackage{algorithm}
\usepackage{algpseudocode}
\DeclareMathAlphabet{\mathcal}{OMS}{cmsy}{m}{n}

\usepackage{theorem}
\theoremstyle{definition}
\newtheorem{defn}{Definition}

\newcolumntype{t}{>{\tt}l}
\newcommand{\tab}{\hspace*{0.75em}}
\newcommand{\blt}{\newmoon}
\newcommand{\crossmark}{\ding{55}}

\newcommand{\includepaper}[3]{\includepdf[pages=-,addtotoc={1,section,1,#2,#3}]{#1}}

\begin{document}

\title{Dynamic Code Coverage with Progressive Detail Levels}
\author{Alexandre Campos Perez}

\thesisdate{18$^{th}$ June, 2012}

\copyrightnotice{Alexandre Campos Perez, 2012}

\supervisor{Supervisor}{Rui Maranhão}{(PhD)}
\supervisor{Co-Supervisor}{André Riboira}{(MSc)}

\committeetext{Approved in oral examination by the committee:}
\committeemember{Chair}{Ademar Manuel Teixeira de Aguiar (PhD)}
\committeemember{External Examiner}{João Alexandre Baptista Vieira Saraiva (PhD)}
\committeemember{Supervisor}{Rui Filipe Maranhão de Abreu (PhD)}
\signature

\logo{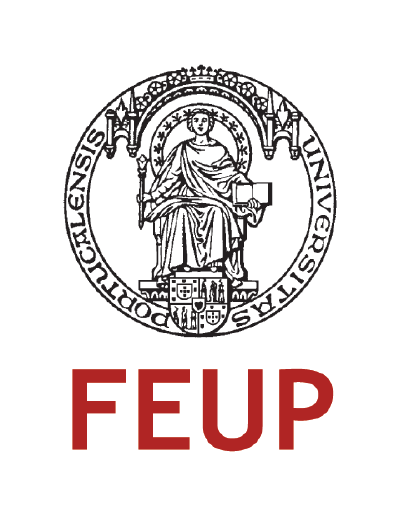}


\begin{Prolog}

\null
\vfill

\noindent {\small This work is financed by the ERDF – European Regional Development Fund through the COMPETE Programme (operational programme for competitiveness) and by National Funds through the FCT - Fundação para a Ciência e a Tecnologia (Portuguese Foundation for Science and Technology) within project PTDC/EIA-CCO/116796/2010.}

\vspace{5mm}

\begin{figure}[ht]
\begin{minipage}[b]{0.35\linewidth}
\centering
\includegraphics[width=\textwidth]{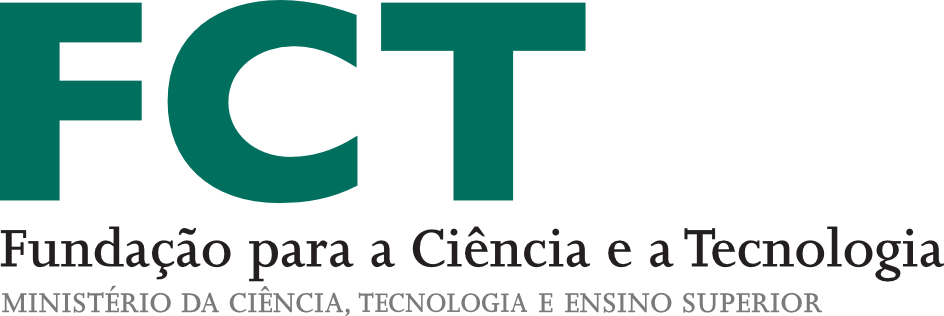}
\end{minipage}
\hfill
\begin{minipage}[b]{0.6\linewidth}
\centering
\includegraphics[width=\textwidth]{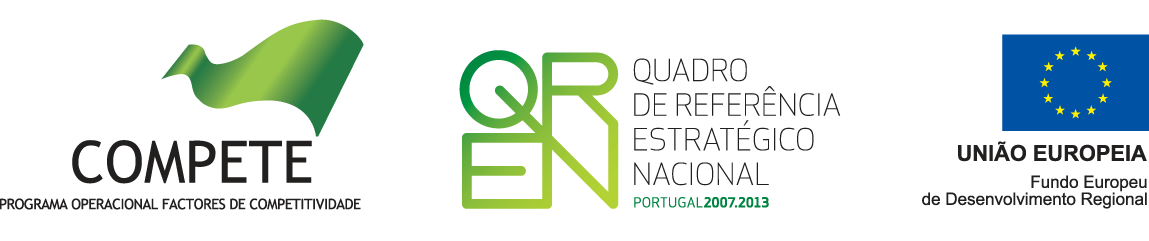}
\end{minipage}
\end{figure}

\chapter*{Resumo}

Hoje em dia, a localização de componentes de software responsáveis por uma avaria é uma das tarefas mais dispendiosas e propensas a erros no processo de desenvolvimento de software.

Para melhorar a eficiência do processo de depuração, algum esforço já foi feito para automaticamente auxiliar a deteção e localização de falhas de software. Isto levou à criação de ferramentas de depuração estatística como Tarantula, Zoltar e GZoltar. Estas ferramentas utilizam informação recolhida a partir de dados de cobertura de código e do resultado das execuções dos casos de teste para retornar uma lista dos locais mais prováveis de conter uma falha.

Estas ferramentas de localização de falhas, apesar de úteis, têm alguns problemas de dimensionamento devido à necessidade de analisar dados de cobertura com granularidade fina. O \emph{overhead} de instrumentação, que em alguns casos pode ser tão elevado como 50\%, é a principal causa para a sua ineficiência.

Esta tese propõe uma nova abordagem para este problema, evitando tanto quanto possível o elevado nível de detalhe de cobertura, mas continuando a utilizar as técnicas comprovadas que as ferramentas de localização de falhas empregam.

Esta abordagem, chamada \acf{DCC}, consiste na utilização de uma instrumentação inicial mais grosseira, obtendo dados de cobertura apenas para componentes de grandes dimensões (\emph{p.e.}, classes). Em seguida, o detalhe da instrumentação de certos componentes é progressivamente aumentado, com base nos resultados intermédios fornecidos pelas mesmas técnicas de localização de falhas utilizadas em ferramentas atuais.

Para avaliar a eficácia da abordagem proposta, foi realizada uma avaliação empírica, injetando falhas em quatro projetos de software. A avaliação empírica demonstra que a abordagem \ac{DCC} reduz o \emph{overhead} de execução que existe nas técnicas atuais de localização de falhas, e também apresenta ao utilizador um relatório de diagnóstico de falha mais conciso. Foi observada uma redução do tempo de execução de 27\% em média e uma redução do tamanho do relatório de diagnóstico de 63\% em média.
\chapter*{Abstract}

Nowadays, locating software components responsible for observed failures is
one of the most expensive and error-prone tasks in the software development
process.

To improve the debugging process efficiency, some effort was already made to
automatically assist the detection and location of software faults. This led
to the creation of statistical debugging tools such as Tarantula, Zoltar and
GZoltar. These tools use information gathered from code coverage data and the
result of test executions to return a list of potential faulty locations.

Although helpful, fault localization tools have some scaling problems because of the 
fine-grained coverage data they need to perform the fault localization 
analysis. Instrumentation overhead, which in some cases can be as high as
50\% is the main cause for their inefficiency.

This thesis proposes a new approach to this problem, avoiding as much as possible
the high level of coverage detail, while still using the proven techniques these
fault localization tools employ.

This approach, named \acf{DCC}, consists of using a coarser 
initial instrumentation, obtaining only coverage traces for large components. Then, the
instrumentation detail of certain components is progressively increased, 
based on the intermediate results provided by the same techniques employed
in current fault localization tools.

To assess the validity of our proposed approach, an empirical evaluation was performed, injecting faults in four real-world software projects. The empirical evaluation demonstrates that the \ac{DCC} approach reduces the execution overhead that exists in spectrum-based fault localization, and even presents a more concise potential fault ranking to the user. We have observed execution time reductions of 27\% on average and diagnostic report size reductions of 63\% on average.

\chapter*{Acknowledgements}

This thesis project would certainly not have been the same without the help of several people and organizations. I would like to take a moment to acknowledge and thank them.

First, I would like to thank Faculdade de Engenharia da Universidade do Porto for providing me with the knowledge that I have gained these last few years.
I would also like to express my utmost gratitude to my supervisors, Prof. Dr. Rui Maranhão and André Riboira for their support, motivation and insight.
Their guidance and feedback helped me greatly throughout this project and I could not imagine having better mentors and advisors for my MSc thesis. It was a pleasure working with them and I hope to be able to work with them again in the future.

A special thanks goes to João Santos, José Carlos de Campos, Nuno Cardoso and Francisco Silva for the all the laughs, support and insightful suggestions during my research at the Software Engineering Laboratory. 

I would also like to give my sincere thanks to my all friends and family for being so supportive and understanding of my absence during stressful periods.

Last, but certainly not least, I would like to thank my parents, Jesus and Maria Filomena, for their unending support throughout my life. 

\vspace{0.5mm}
\flushleft{Porto, 18$^{th}$ June, 2012}

\vspace{10mm}
\flushleft{Alexandre Perez}
  \cleardoublepage
  \pdfbookmark[0]{Table of Contents}{contents}
  \tableofcontents
  \cleardoublepage
  \pdfbookmark[0]{List of Figures}{figures}
  \listoffigures
  \cleardoublepage
  \pdfbookmark[0]{List of Tables}{tables}
  \listoftables
  \cleardoublepage
  \pdfbookmark[0]{List of Algorithms}{algorithms}
  \listofalgorithms
  \cleardoublepage
\chapter*{Abbreviations}
\chaptermark{ABBREVIATIONS}

\acrodefplural{LOC}[LOCs]{Lines Of Code}

\begin{flushleft}
	\begin{acronym}
		\acro{DCC}{Dynamic Code Coverage}
		\acro{IDE}{Integrated Development Environment}
		\acro{JVM}{Java Virtual Machine}
		\acro{LOC}{Line Of Code}
		\acro{MBD}{Model-Based Diagnosis}
		\acro{MBSD}{Model-Based Software Debugging}
		\acro{SDT}{Statistical Debugging Tool}
		\acro{SFL}{Spectrum-based Fault Localization}
		\acro{SUT}{System Under Test}
	\end{acronym}
\end{flushleft}
\end{Prolog}

\StartBody

\acresetall
\chapter{Introduction} 
\label{cha:introduction}


The software development process generally follows four main phases: a requirements and design phase,
after that an implementation phase, followed by a testing phase, and finally the release. All these steps
can be defined, and estimated, with a high degree of certainty.

\begin{figure}[h]
\begin{center}
	\includegraphics[width=0.5\textwidth]{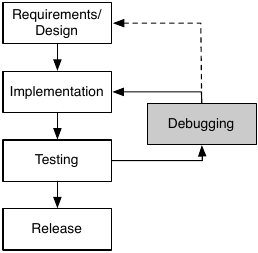}
	\caption{Software development process.}
	\label{fig:software-process}
\end{center}
\end{figure}

However, in most, if not all, software projects, some tests fail. Because of that, cycles are introduced
in the process, in which another task has to be performed -- the so-called debugging phase (see
Figure~\ref{fig:software-process}).

Most of the time, the debugging phase consists of changing the implementation so that faults are eliminated. However, in some cases, the system design itself can be at fault, and has to be modified. In this thesis, we will focus on the software implementation debugging.

The debugging phase consumes a huge amount of a project's resources~\cite{Tassey2002,Hailpern2002}. Furthermore, one
cannot estimate with a high degree of certainty the cost of this phase (in terms of time and also money). 
For this reason, it is important to minimize the impact that debugging has in the development process. 

Software debugging tools and methodologies almost always existed, but they were fairly ineffective and \emph{ad-hoc}. Currently, there are some tools that automate this process by returning the most likely locations of 
containing a fault. However, these tools do not scale because they need to instrument code at a fine-grained detail
level.  

This thesis' main goal is to improve the efficiency of automated debugging techniques so that developers 
spend less time locating faults and thus minimizing the cost of this cumbersome phase.

\section{Context} 
\label{sec:context}

\begin{figure}[h]
\begin{center}
	\includegraphics[width=0.95\textwidth]{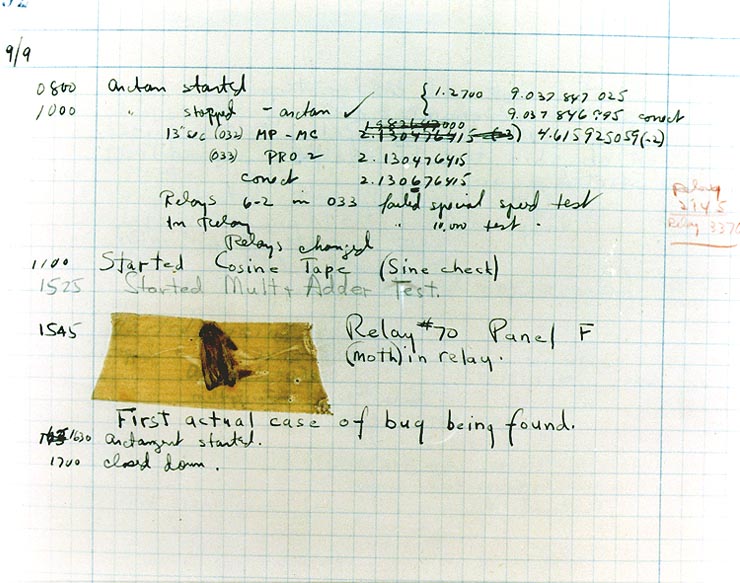}
	\caption{First actual case of bug being found.}
	\label{fig:first-bug}
\end{center}
\end{figure}

In 1947, the Harvard Mark II was being tested by Grace Murray Hopper and her associates when the machine
suddenly stopped. Upon inspection, the error was traced to a dead moth that was trapped in a relay and 
had shorted out some of the circuits. The insect was removed and taped to the machine's logbook (see 
Figure~\ref{fig:first-bug})~\cite{Kidwell1947}. This incident coined the use of the terms ``bug'', ``debug''
and ``debugging'' in the field of computer science.

In software development, a large amount of resources is spent in the debugging phase. It is estimated that testing and debugging activities can easily range from 50 to 75 percent of the total development cost~\cite{Hailpern2002}.
This is due to the fact that the process of detecting, locating and fixing faults in the source code, is not trivial and rather error-prone. Even experienced developers are wrong almost 90\% of the time in their initial guess while trying to identify the cause of a behavior that deviates from the intended one \cite{Ko2008}. 

If this debugging task is not thoroughly conducted, even bigger costs may arise. In fact, a landmark study performed in 2002 indicated that software defects constitute an annual \$60 billion cost to the US economy alone~\cite{Tassey2002}.


The first debugging techniques to be used consisted of utilizing prints and stack traces, assertions, breakpoints, coverage information and profiling. These techniques, which sometimes are called traditional debugging techniques, are quite ineffective by themselves and rather \emph{ad-hoc} in nature. 

Traditional debugging techniques have several limitations. They rely heavily on the developer's intuition and suffer from a considerable execution overhead, since many combinations of program executions have to be examined. Furthermore, in most cases, developers only use failing test cases to diagnose a certain defect, disregarding useful information about passing test cases. Traditional debugging also requires developers to have a comprehensive knowledge of the program under test. 

In order to improve the debugging efficiency, this process needs to be automated. Some effort was already made
to automatically assist the detecting and locating steps of the software debugging phase. This led to the creation of automatic fault localization
tools, namely Zoltar \cite{Janssen2009} and Tarantula \cite{Jones2002}. The tools instrument the source code to
obtain code coverage traces for each execution (also known as program spectra), which are then analyzed to return a list of potential faulty locations.

To improve the exploration and intuitiveness of that potential faulty locations list, an Eclipse\footnote{Eclipse integrated development environment -- \url{http://www.eclipse.org/}} plugin was also developed --
GZoltar \cite{Riboira2010}. This tool provides fault localization functionality to an \ac{IDE}, with several interactive visualization options, as well as testing selection and prioritization functionalities.


\section{Concepts and Definitions} 
\label{sec:concepts_and_definitions}

Throughout this thesis, we use the following terminology~\cite{avizienis04concepts}:

\begin{itemize}
\item A \textit{failure} is an event that occurs when delivered service deviates from correct service.

\item An \textit{error} is a system state that may cause a failure.

\item A \textit{fault} (defect/bug) is the cause of an error in the system.
\end{itemize}

In this thesis, we apply this terminology to software programs, where faults are bugs in the program code. Failures and errors are symptoms caused by faults in the program. The purpose of fault localization is to pinpoint the root cause of observed symptoms.

\begin{defn}
A software program $\Pi$ is formed by a sequence of one or more $M$ statements. 
\end{defn}

Given its dynamic nature, central to the fault localization technique considered in this thesis is the existence of a test suite. 

\begin{defn}
A test suite $T = \{t_1, \ldots, t_N\}$ is a collection of test cases that are intended to test whether the program follows the specified set of requirements. The cardinality of $T$ is the number of test cases in the set $|T| = N$.
\end{defn}

\begin{defn}
A test case $t$ is a $(i,o)$ tuple, where $i$ is a collection of input settings or variables for determining whether a software system works as expected or not, and $o$ is the expected output. If $\Pi(i) = o$ the test case passes, otherwise fails.
\end{defn}


\section{Motivation} 
\label{sec:motivation}

It is essential to find ways to minimize the software testing and debugging impact on a project's resources. However, it is imperative that the software quality (\emph{i.e.}, correctness) is not compromised. While some defects can be tolerated by users (or even not perceived at all), others may cause severe financial and/or life-threatening consequences. Examples of well-known drastic consequences caused by software defects are:

\begin{itemize}
	\item The software malfunction of the rocket Ariane 5, which caused it to disintegrate 37 seconds after its launch~\cite{Lions1996,Dowson1997};
	\item The crash of a British Royal Air Force Chinook due to a software defect in the helicopter's engine control computer, killing 29 people~\cite{Rogerson2002}.
\end{itemize}

For this reason, automatic debugging tools are essential to aid developers in maintaining their software project's quality. Nowadays, automatic fault localization techniques can aid developers/testers in pinpointing the root cause of software failures, and thereby reducing the debugging effort. Amongst the most diagnostic-effective techniques is \ac{SFL}. 

\ac{SFL} is a statistical technique that uses abstraction of program traces (also known as program spectra) to correlate software component (e.g., statements, methods, classes) activity with program failures~\cite{Abreu2009a,sober,crosstab}. As SFL is typically used to aid developers in identifying what is the root cause of observed failures, it is used with high level of granularity (\emph{i.e.}, statement level). 

Statistical approaches to debugging are very attractive because of the relatively small overhead with respect to CPU time and memory requirement~\cite{Abreu2009a,ase09}. However, gathering the input information, per test case, to yield the diagnostic ranking may still impose a considerable (CPU time) overhead. This is particularly the case for resource constrained environments, such as embedded systems.

As said before, typically, \ac{SFL} is used at development-time at a statement level granularity (since debugging requires to locate the faulty statement). But not all components need to be inspected at such fine grain granularity. In fact, components that are unlikely to be faulty may not need to be inspected. This way, by removing instrumentation from unlikely locations, more projects will be suitable to be debugged with these fault localization techniques, since the imposed overhead will be reduced.


\section{Research Question} 
\label{sec:research_questions}

The main research question that we are trying to answer with this work is the following:

\begin{itemize}
	\item {How can a fault localization approach that instruments less software components obtain similar diagnostic results when compared with \ac{SFL}, while reducing execution overhead?}
\end{itemize}

The main objective of this thesis is to devise a fault localization approach that is comparable to \ac{SFL} in terms of diagnostic results, while trying to be more lightweight regarding the instrumentation used to obtain the characterization of the program executions (\emph{i.e.}, the program spectra).

This way, bigger projects will be able to use these debugging methodologies than before (especially resource constrained projects). It is also expected that due to this instrumentation decrease, the fault localization process will be shorter in terms of execution time.


\section{A Dynamic Code Coverage Approach} 
\label{sec:dynamic_code_coverage_approach}

This thesis proposes a technique, coined \ac{DCC}.

\begin{figure}[H]
\begin{center}
	\includegraphics[width=0.7\textwidth]{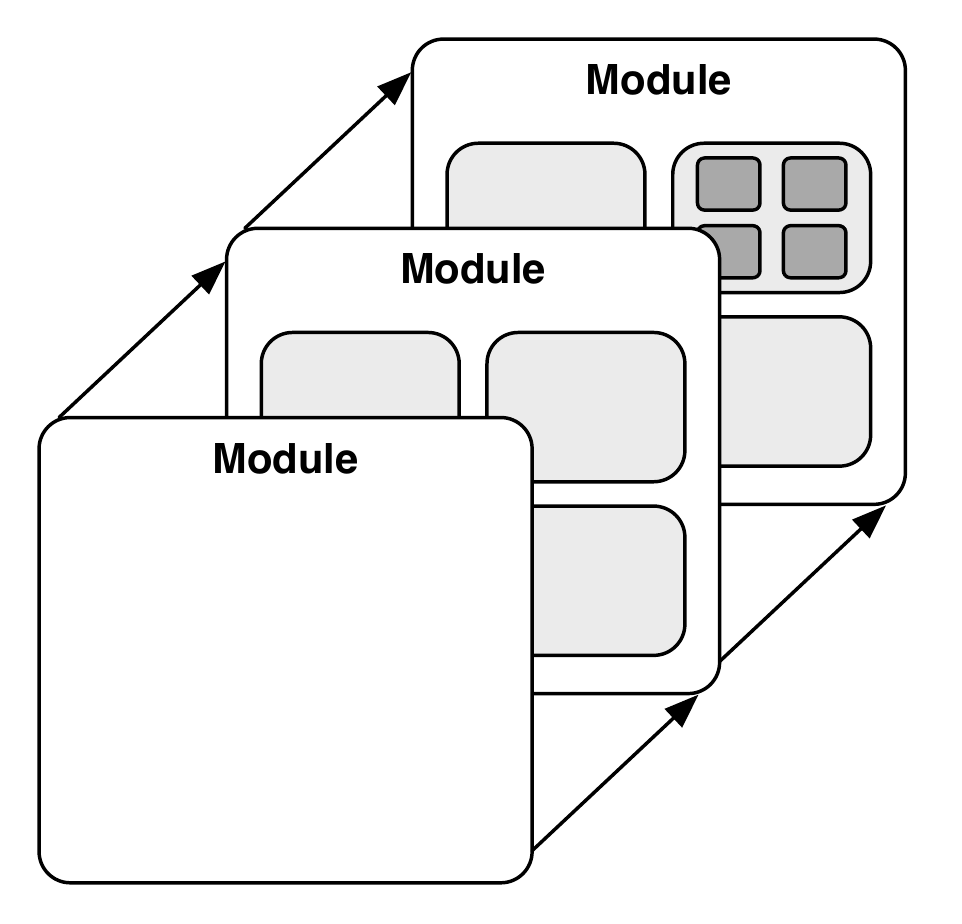}
	\caption{Progressive detail of a component.}
	\label{fig:hierarchy}
\end{center}
\end{figure}

This technique automatically adjusts the granularity per software component. First, our approach instruments the source code using a coarse granularity (e.g., package level in Java) and the fault localization is executed. Then, it is decided which components are \textit{expanded} based on the output of the fault localization technique. With expanding we mean changing the granularity of the instrumentation to the next detail level (e.g., in Java, for instance, instrument classes, then methods, and finally statements), behaving like the diagram in Figure~\ref{fig:hierarchy}. This expansion can be done in different ways, either selecting the components whose fault coefficient is above a certain threshold, or selecting the first ranked components, according to a set percentage.

\section{Document Structure} 
\label{sec:document_structure}

This document will be structured as follows:

Chapter~\ref{cha:sota} contains the state of the art in this project's field. Traditional and statistical debugging techniques and tools, and model-based reasoning approaches to debugging are presented in this chapter.

Chapter~\ref{cha:dcc} details the \ac{DCC} algorithm. A motivational example is presented to show the inefficiencies of current approaches, followed by the \ac{DCC} algorithm description and its main advantages and shortcomings.

Chapter~\ref{cha:tooling} presents the tool chosen to host the \ac{DCC} prototype -- GZoltar -- as well as some modifications that have been made to this tool.

Chapter~\ref{cha:empirical_evaluation} presents our empirical evaluation setup and findings.

Chapter~\ref{cha:conclusions} presents some conclusions about the work and describes the main contributions of this thesis. After that, some future work challenges are presented.

Lastly, Appendix~\ref{cha:publications} contains the accepted scientific publications during this work, submitted to \emph{IJUP'12}\footnote{The 5th Meeting of Young Researchers of University of Porto, 2012} and \emph{ASE'12}\footnote{The 27th IEEE/ACM International Conference on Automated Software Engineering, 2012}. A publication currently being prepared for submission into \emph{ICST'13}\footnote{The 6th IEEE International Conference on Software Testing, Verification and Validation, 2013} is also included.

Note that most of the work presented in this thesis has been submitted for publication at \emph{ICST'13}.



\acresetall
\chapter{State of the art} 
\label{cha:sota}

In software development, a large amount of resources is spent in debugging~\cite{Tassey2002,Hailpern2002}. This 
process consists of three major phases:
\begin{itemize}
	\item{\textbf{Detecting} a fault in a program's behavior.}
	\item{\textbf{Locating} said fault in the source code.}
	\item{\textbf{Fixing} the code to eliminate the fault.} 
\end{itemize}

This not trivial and rather error-prone. For this reason, many tools were created to assist developers in 
this process, each tool having its advantages and its disadvantages. This chapter presents the state of the 
art of debugging techniques.

\section{Traditional Debugging} 
\label{sec:traditional_debugging}

In this section, some traditional debugging techniques and tools will be described, namely assertions, breakpoints,
profiling and code coverage.

\subsection{Assertions} 
\label{sub:assertions}

Assertions are formal constraints that the developer can use to specify \emph{what} the system is supposed to do
(rather than \emph{how})~\cite{Rosenblum1995}.

These constructs are generally predefined macros that expand into an \emph{if} statement that aborts the execution
if the expression inside the assertion evaluates to false. 

Assertions can be seen, then, as permanent defense mechanisms for runtime fault detection.


\subsection{Breakpoints} 
\label{sub:breakpoints}

A breakpoint specifies that the control of a program execution should transfer to the user when a specified instruction
is reached~\cite{Corliss2005}. The execution is stopped and the user can inspect and manipulate its state (\emph{e.g.} 
the user can read and change variable values). It is also possible to perform a step-by-step execution after the 
breakpoint. This is particularly useful to observe a bug as it develops, and to trace it to its origin.

There are other types of breakpoints, namely data breakpoints and conditional breakpoints. Data breakpoints (also called
watchpoints~\cite{Stallman2002}) transfer control to the user when the value of a expression changes. This expression may 
be a value of a  variable, or multiple variables combined by operators (\emph{e.g.}: \texttt{a + b}). Conditional 
breakpoints only stop the execution if a certain user-specified predicate is true, thus reducing the frequency of
user-application interaction.

Breakpoints are available in most modern \acp{IDE}.


\subsection{Profiling} 
\label{sub:profiling}


Profiling is a dynamic analysis that gathers some metrics from the execution of a program, such as memory usage and
frequency and duration of function calls. Profiling's main use is to aid program optimization, but it is also useful
for debugging purposes, such as:

\begin{itemize}
	\item Knowing if functions are being called more or less often than expected;
	\item Finding if certain portions of code execute slower than expected or if they contain memory leaks;
	\item Investigating the behavior of lazy evaluation strategies.
\end{itemize}

Known profiling tools include GNU's gprof~\cite{Graham1982} and the Eclipse plugin 
TPTP\footnote{Eclipse Test \& Performance Tools Platform Project -- \url{http://www.eclipse.org/tptp/}}.



\subsection{Code Coverage} 
\label{sub:code_coverage}

Code coverage is an analysis method that determines which parts of the \ac{SUT} have been executed (covered) during
a system test run~\cite{Graham2006}.

Using code coverage in conjunction with tests, it is possible to see which 
\acp{LOC}, methods or classes were covered in a specific test (depending on the set level of detail). With this
information, it is possible to identify which components were involved in a system failure, narrowing the search
for the faulty component that made the test fail.

Table~\ref{tab:coverage-programs} presents a non-exhaustive list of code coverage tools available in the market, with
the languages they support and the detail levels they provide (adapted from~\cite{Yang2006}). It is worth to note that we have limited the scope of our research to code coverage tools for imperative and object oriented languages. This way, a comparison using coverage measurement detail levels can be established. 

\begin{table}[H]
\begin{center}
	\scalebox{0.85}{
\begin{tabular}{|l|p{6.5cm}|p{2cm}|c|c|c|c|}
	\hline
																& \bf{Website}	& \bf{Language(s)} 							& \bf{Line}	& \bf{Decision}	& \bf{Method}	& \bf{Class}	\\ \hline
	Agitar  &\footnotesize{\url{http://www.agitar.com/}}					& Java 													& \checkmark	& \checkmark	& \checkmark	& \checkmark	\\ \hline
	Bullseye  &\footnotesize{\url{http://www.bullseye.com/}}		& C/C++ 												& 						& \checkmark	& \checkmark	& 						\\ \hline
	Clover &\footnotesize{\url{http://www.atlassian.com/software/clover/}}				& Java, .NET 										& \checkmark	& \checkmark	& \checkmark	& 						\\ \hline
	Cobertura &\footnotesize{\url{http://cobertura.sourceforge.net/}}		& Java 													& \checkmark	& \checkmark	& 						& 						\\ \hline
	Dynamic &\footnotesize{\url{http://www.dynamic-memory.com/}}				& C/C++ 												& \checkmark	& \checkmark	& \checkmark	& 						\\ \hline
	EclEmma &\footnotesize{\url{http://www.eclemma.org/}}				& Java 													& \checkmark	& \checkmark	& \checkmark	& \checkmark	\\ \hline
	gcov &\footnotesize{\url{http://gcc.gnu.org/onlinedocs/gcc/Gcov.html}}						& C/C++ 												& \checkmark	& 						& 						& 						\\ \hline
	Insure++  &\footnotesize{\url{http://www.parasoft.com/jsp/products/insure.jsp}}				& C/C++ 												& \checkmark	& 						& 						& 						\\ \hline
	JCover  &\footnotesize{\url{http://www.mmsindia.com/JCover.html}} 				& Java 													& \checkmark	& \checkmark	& \checkmark	& \checkmark	\\ \hline
	JTest  &\footnotesize{\url{http://www.parasoft.com/jsp/products/jtest.jsp}} 					& Java, .NET 										& \checkmark	& \checkmark	& 						&							\\ \hline
	PurifyPlus  &\footnotesize{\url{http://www.ibm.com/software/awdtools/purifyplus/}} & C/C++, Java, .NET 						& \checkmark	& 						& \checkmark	& 						\\ \hline
	SD  &\footnotesize{\url{http://www.semdesigns.com/}} 								& C/C++, Java, C\#, PHP, COBOL 	& \checkmark	& \checkmark	& \checkmark	& \checkmark	\\ \hline
	TCAT  &\footnotesize{\url{http://www.soft.com/Products/Coverage/}} 						& C/C++, Java 									& \checkmark	& \checkmark	& \checkmark	& \checkmark	\\ \hline
\end{tabular}}
\end{center}
\caption{Code Coverage tools comparison.}
\label{tab:coverage-programs}
\end{table}

In order to obtain information about what components were covered in each run, these Code Coverage tools
have to instrument the system code. This instrumentation will monitor each component and register if the
they were executed.

\begin{figure}[h]
\begin{center}
	\includegraphics[width=0.95\textwidth]{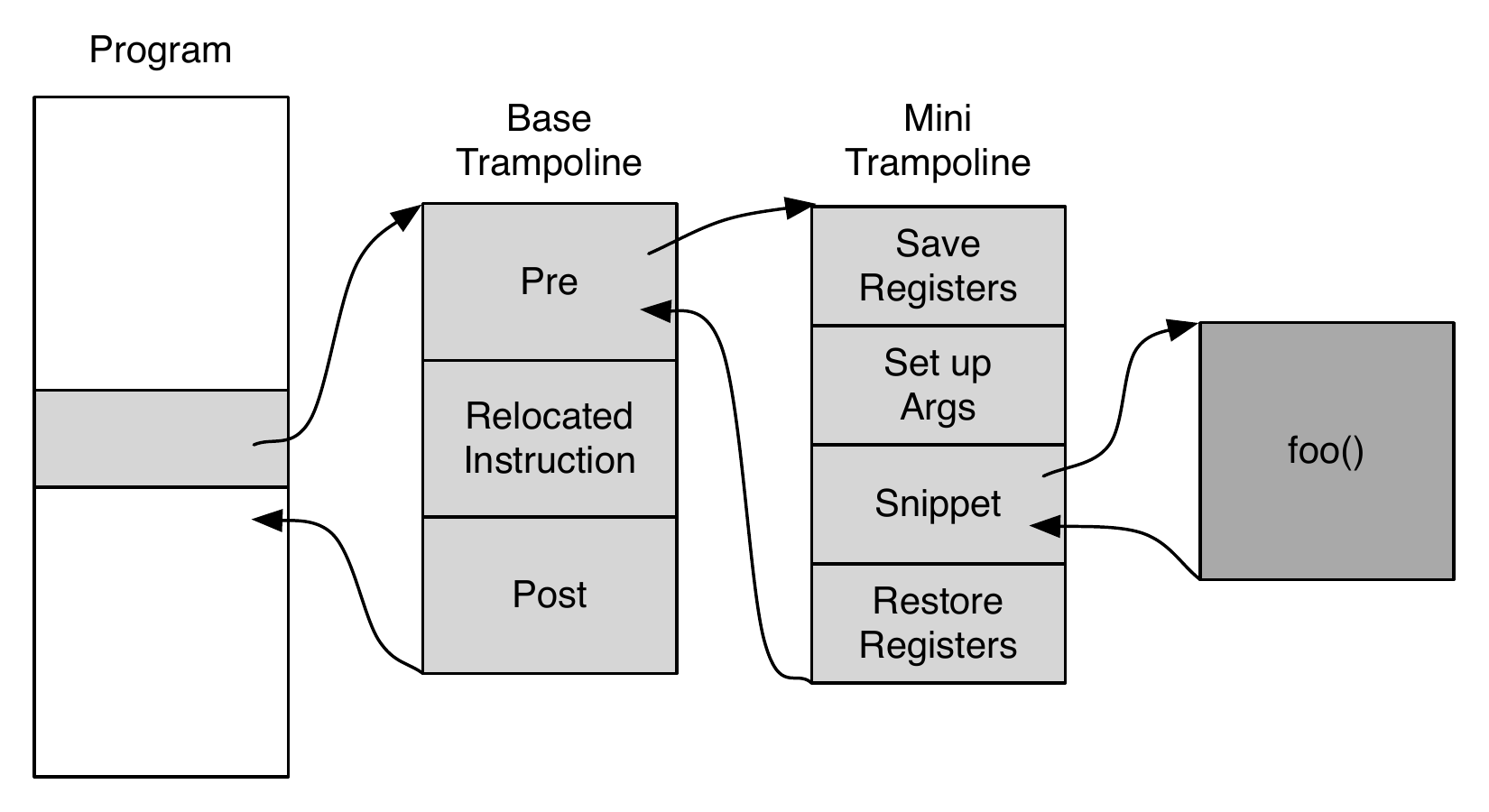}
	\caption{Instrumentation Code Insertion~\cite{Tikir2002}.}
	\label{fig:instrumentation-program}
\end{center}
\end{figure}

Instrumentation code, as seen in Figure~\ref{fig:instrumentation-program}, relies in a series of trampolines
to a desired function \texttt{foo()} before the desired instructions. In the case of Cove Coverage
tools, \texttt{foo()} will register that the instruction was touched by the execution.

This instrumentation introduces overhead to the system's execution during the testing and debugging phases. According to Yang, et al~\cite{Yang2006},
Code Coverage tools, which have to instrument at a \ac{LOC} level, introduce execution speed overheads of up to 50\%.


\section{Statistical Debugging} 
\label{sec:statistical_debugging}

\acp{SDT} use statistical techniques to calculate the probability of a certain software component of the \ac{SUT} 
containing faults. The most effective statistical technique is \ac{SFL}~\cite{Abreu2009}. 

\ac{SFL} exploits information from passed and failed system runs. A passed run is a program execution that is 
completed correctly, and a failed run is an execution where an error was detected~\cite{Janssen2009}. The criteria 
for determining if a run has passed or failed can be from a variety of different sources, namely test case results 
and program assertions, among others. The information gathered from these runs is their code coverage (also called 
program spectra~\cite{Abreu2006}). 

A program spectra is a characterization of a program's execution on a dataset~\cite{Reps1997}. This collection of data, gathered at runtime, provides a view on the dynamic behavior of a program. The data consists of counters or flags for each software component. Different program spectra exist~\cite{Harrold2000}, such as path-hit spectra, data-dependence-hit spectra, and block-hit spectra.

As explained in section~\ref{sub:code_coverage}, in order to obtain information about which components were covered in each execution, the program's source code needs to be instrumented, similarly to code coverage tools~\cite{Yang2006}. This instrumentation will monitor each component and register those that were executed. Components can be of several detail granularities, such as classes, methods, and lines of code.

The hit spectra of $N$ runs constitutes a binary $N \times M$ matrix $A$, where $M$ corresponds to the instrumented 
components of the program. Information of passed and failed runs is gathered in a $N$-length vector $e$, called the
error vector. The pair $(A,e)$ serves as input for the \ac{SFL} technique, as seen in Figure~\ref{fig:sfl-matrix}. 

\begin{figure}[h]
$$
  \text{$N$ spectra }
  \stackrel{\mbox{$M$ components}}{%
    \begin{bmatrix}
    a_{11} & a_{12} & \cdots & a_{1M} \\
    a_{21} & a_{22} & \cdots & a_{2M} \\
    \vdots & \vdots & \ddots & \vdots \\
    a_{N1} & a_{N2} & \cdots & a_{NM}
    \end{bmatrix}%
  }\ \ \
  \stackrel{\stackrel{\mbox{error}}{\mbox{detection}}}{%
    \begin{bmatrix}
    e_1 \\
    e_2 \\
    \vdots \\
    e_N
    \end{bmatrix}%
   }
$$
\caption{Input to \ac{SFL}~\cite{Janssen2009}.}
\label{fig:sfl-matrix}
\end{figure}

With this input, fault localization consists in identifying what columns of the matrix $A$ resemble the vector
$e$ the most. For that, several different similarity coefficients can be used~\cite{Jain1988}. One of them is the 
Ochiai coefficient~\cite{Abreu2007}, used in the molecular biology domain. Ochiai is defined as follows:

\begin{equation}\label{eq:ochiai}
	s_O(j) = \frac{n_{11}(j)}{\sqrt{(n_{11}(j)+n_{01}(j))\times(n_{11}(j)+n_{10}(j))}}
\end{equation}

\noindent where $n_{pq}(j)$ is the number of runs in which the component $j$ has been touched during execution ($p = 1$) or not touched during execution ($p=0$), and where the runs failed ($q = 1$) or passed ($q = 0$). For instance, $n_{11}(j)$ counts the number of times component $j$ has been involved in failed executions, whereas $n_{10}(j)$ counts the number of times component $j$ has been involved in passed executions. Formally, $n_{pq}(j)$ is defined as
\begin{equation}
 n_{pq}(j) = |\{i\mid a_{ij}=p \wedge e_i=q\}|
\end{equation}

\begin{table}
\begin{center}

\begin{tabular}{tt|c|c|c|c|c|c|c}
	\hline
	&mid() \{ 
	& \multicolumn{6}{c|}{Runs} & \\ \cline{3-8}
	&\tab int x,y,z,m; 
	& 1 & 2 & 3 & 4 & 5 & 6 & Coef. \\ \hline
	1:&\tab read("Enter 3 numbers:",x,y,z); 
	& \blt & \blt & \blt & \blt & \blt & \blt & 0.41 \\ \hline
  2:&\tab m = z;  
	& \blt & \blt & \blt & \blt & \blt & \blt & 0.41 \\ \hline
	3:&\tab if (y<z) \{  
	& \blt & \blt & \blt & \blt & \blt & \blt & 0.41 \\ \hline
  4:&\tab \tab if (x<y)  
	& \blt & \blt &  &  & \blt & \blt & 0.50 \\ \hline
	5:&\tab \tab \tab m = y;  
	&  & \blt &  &  &  &  & 0.0 \\ \hline
	6:&\tab \tab else if (x<z)  
	& \blt &  &  &  & \blt & \blt & 0.58 \\ \hline
	7:&\tab \tab \tab m = y; //BUG
	& \blt &  &  &  & \blt &  & 0.71 \\ \hline
	8:&\tab \} else \{  
	&  &  & \blt & \blt &  &  & 0.0\\ \hline
	9:&\tab \tab if (x>y)  
	&  &  & \blt & \blt &  &  & 0.0 \\ \hline
	10:&\tab \tab \tab m = y;  
	&  &  & \blt &  &  &  & 0.0 \\ \hline
	11:&\tab \tab else if (x>z)  
	&  &  &  & \blt &  &  & 0.0 \\ \hline
	12:&\tab \tab \tab m = x;  
	&  &  &  &  &  &  & 0.0 \\ \hline
	13:&\tab \}  
	&  &  &  &  &  &  & 0.0 \\ \hline
	14:&\tab print("Middle number is:",m);  
	& \blt & \blt & \blt & \blt & \blt & \blt & 0.41 \\ \hline
	&\} \hfill \textrm{Pass/fail status:}
	& \checkmark & \checkmark & \checkmark & \checkmark & \crossmark & \checkmark & \\ \hline
\end{tabular}
\end{center}
\caption{Example of \ac{SFL} technique with Ochiai coefficient.}
\label{tab:sfl-example}
\end{table}

In Table~\ref{tab:sfl-example} it is shown an example of the \ac{SFL} technique, using the Ochiai coefficient 
(adapted from~\cite{Jones2005}). To improve this example's legibility, the coverage matrix and the error detection vector were transposed. In this example, the \ac{SUT} is a function named \texttt{mid()} that reads 
three integer numbers and prints the median value. This program contains a fault on line 7 -- it should 
read \texttt{m = x;}. 

Six test cases were run, and their coverage information for each \ac{LOC} can be seen to 
the right. At the bottom there is also the pass/fail status for each run -- which corresponds to the error detection vector $e$. Then, the similarity coefficient was calculated for each line using the Ochiai coefficient~(\ref{eq:ochiai}). These 
results represent the likelihood of a certain line containing a fault. The bigger the coefficient, the more likely 
it is of a line containing a fault. Therefore, these coefficients can be ranked to form an ordered list of the probable 
faulty locations.

In this specific example, the highest coefficient is in line 7 -- the faulty \ac{LOC}. The \ac{SFL} technique has
successfully performed the fault localization.

\subsection{Tarantula} 
\label{sub:tarantula}

Tarantula\footnote{Tarantula - Fault Localization via Visualization -- \url{http://pleuma.cc.gatech.edu/aristotle/Tools/tarantula/}}~\cite{Jones2002} is a visual debugging system that is used to debug projects written in the C language. This tool is being developed at Georgia Tech and uses \ac{SFL} for the fault localization.

\begin{figure}[H]
\begin{center}
	\includegraphics[width=1\textwidth]{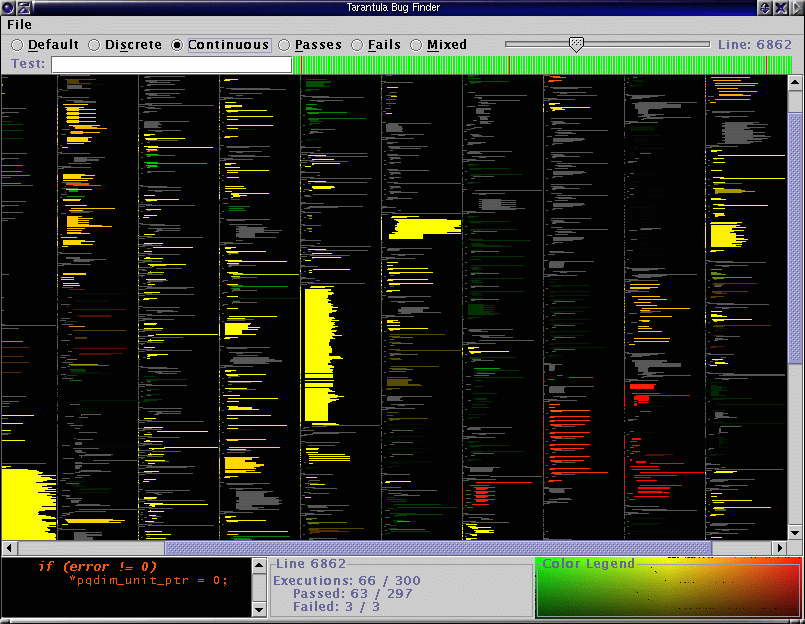}
	\caption{Tarantula interface.}
	\label{fig:tarantula}
\end{center}
\end{figure}

Tarantula runs test suites against the \ac{SUT} and displays the calculated probability of each \ac{LOC} by highlighting them 
accordingly -- varying from red (maximum failure probability) to green (minimum failure probability). In
Figure~\ref{fig:tarantula} it is shown the tool's interface. Tarantula has the ability to analyze the whole system at
once, which is convenient for debugging large projects, as well as gathering management metrics about the project.  

For the similarity coefficient, Tarantula uses the following coefficient~\cite{Abreu2007,Jones2005}:

\begin{equation}\label{eq:tarantula}
	s_T(j) = \frac{\frac{n_{11}(j)}{n_{11}(j)+n_{01}(j)}}{ \frac{n_{11}(j)}{n_{11}(j)+n_{01}(j)} + \frac{n_{10}(j)}{n_{10}(j)+n_{00}(j)}}
\end{equation}

However, studies~\cite{Abreu2007} have shown that this coefficient under-performs Ochiai~(\ref{eq:ochiai}). Furthermore, Tarantula is currently not available for download.


\subsection{Zoltar} 
\label{sub:zoltar}

\begin{figure}[H]
\begin{center}
	\includegraphics[width=1\textwidth]{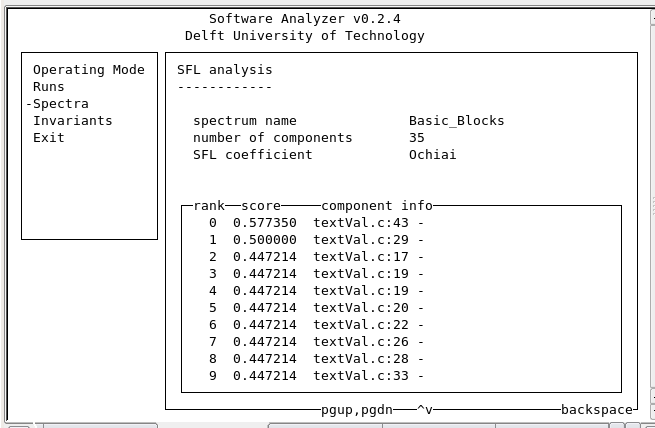}
	\caption{Zoltar interface~\cite{Janssen2009}.}
	\label{fig:zoltar}
\end{center}
\end{figure}

\begin{figure}[H]
\begin{center}
	\includegraphics[width=0.9\textwidth]{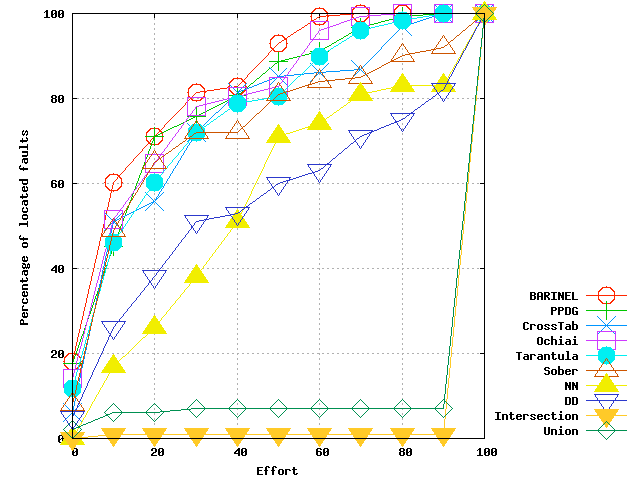}
	\caption{\ac{SFL}'s similarity coefficients performance comparison~\cite{Abreu2009}.}
	\label{fig:sfl-performance}
\end{center}
\end{figure}

Zoltar~\cite{Janssen2009} is a tool that implements fault localization in C/C++ projects (see Figure~\ref{fig:zoltar}). This tool currently presents superior performance compared with similar tools, and has implemented several algorithms, namely Barinel, Ochiai and Tarantula~\cite{Abreu2009}. Figure~\ref{fig:sfl-performance} shows the comparison between several Zoltar algorithms.

Zoltar was developed at Delft University of Technology (TUDelft), and was the base of Rui Abreu's PhD thesis~\cite{Abreu2009}. This tool had substantial academic recognition, and won the \emph{Best Demo Award} prize at \emph{The 24$^{th}$ IEEE/ACM International Conference on Automated Software Engineering (ASE'09)} with the publication \emph{Zoltar: A Toolset for Automatic Fault Localization}~\cite{Janssen2009b}.

Zoltar has already proven to be effective in real world uses, from the development of embedded devices, such as TV sets, to sizable internal software projects.

\subsection{EzUnit} 
\label{sub:ezunit}

EzUnit\footnote{EzUnit -- Easing the Debugging of Unit Test Failures -- \url{http://www.fernuni-hagen.de/ps/prjs/EzUnit4/}} is a statistical debugging tool under development at University of Hagen. This tool is integrated into the Eclipse \ac{IDE} as a plugin, and is used to debug Java projects that use JUnit test cases.

\begin{figure}[H]
\begin{center}
	\includegraphics[width=1\textwidth]{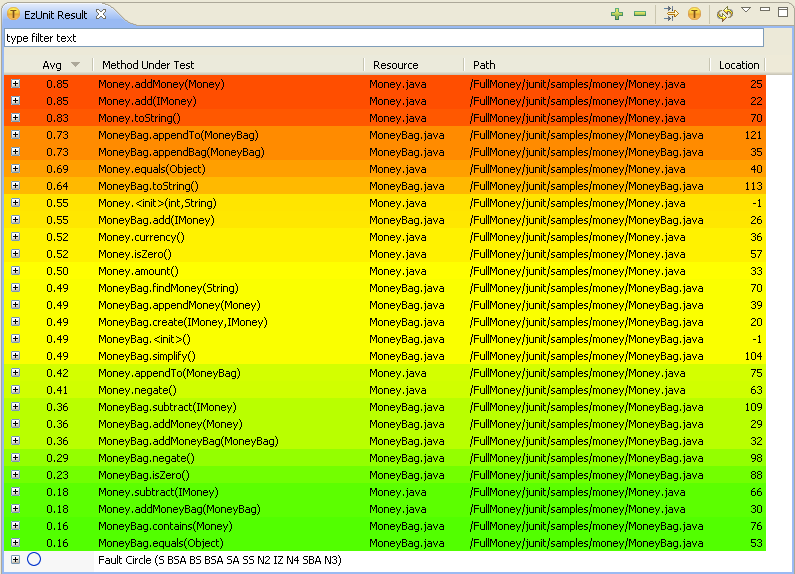}
	\caption{EzUnit interface.}
	\label{fig:ezunit1}
\end{center}
\end{figure}

\begin{figure}[H]
\begin{center}
	\includegraphics[width=0.8\textwidth]{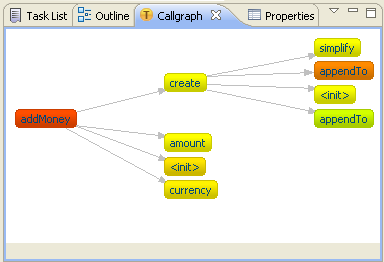}
	\caption{EzUnit call graph.}
	\label{fig:ezunit2}
\end{center}
\end{figure}

After performing the fault localization, EzUnit displays a list of the code blocks ranked by their failure probability in a separate view in the Eclipse \ac{IDE} (see Figure~\ref{fig:ezunit1}). Each line of the list is highlighted with a color representing the failure probability, ranging from red (corresponding to the code blocks that are most likely to contain a failure) to green (least likely). EzUnit also provides a call-graph view of a certain test (see Figure~\ref{fig:ezunit2}). Each node in the graph corresponds to a code block, and has the same coloring scheme as the failure probability list view.

\subsection{GZoltar} 
\label{sub:gzoltar}

\begin{figure}[H]
\begin{center}
	\includegraphics[width=1\textwidth]{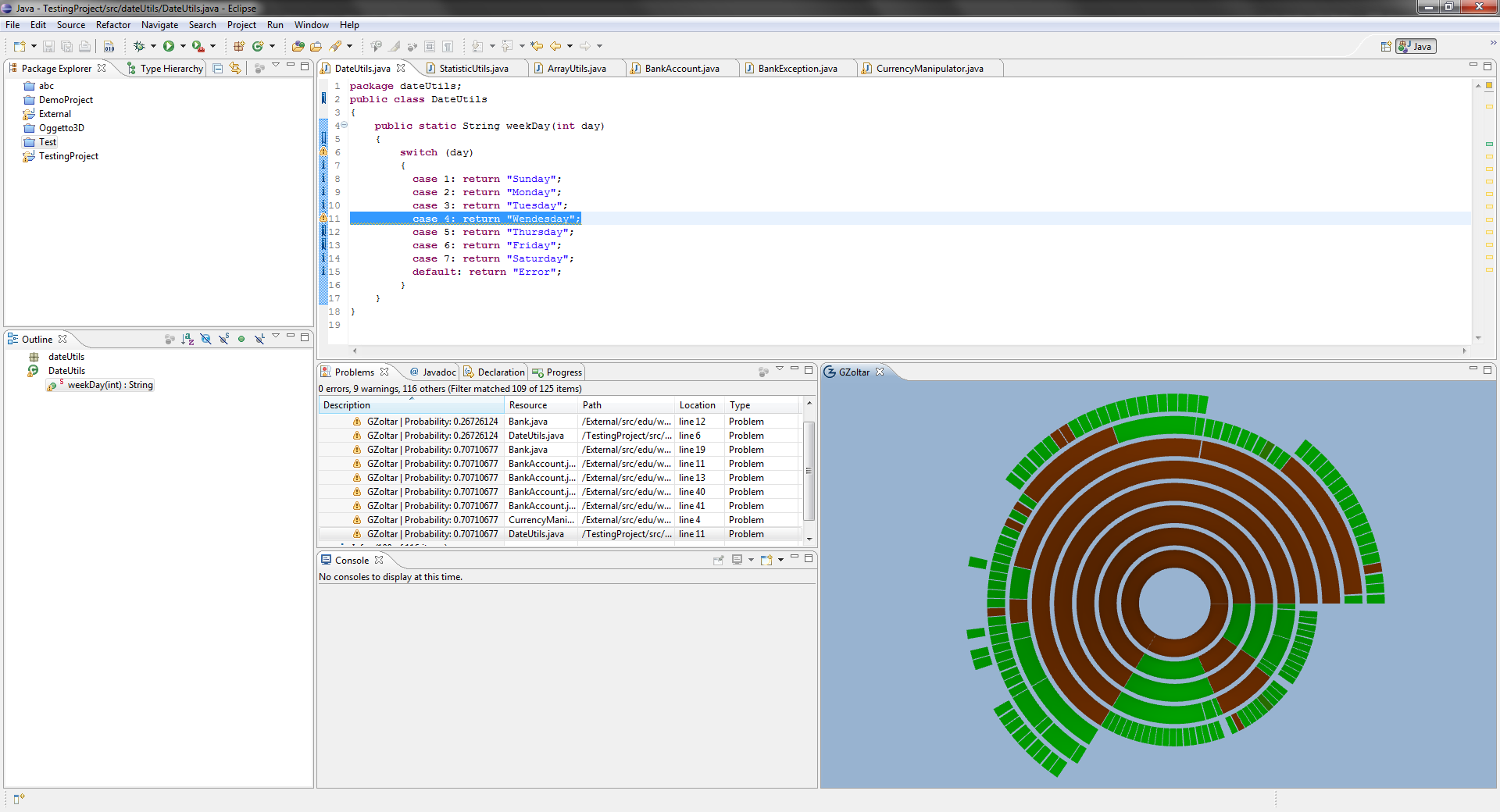}
	\caption{GZoltar interface~\cite{Riboira2011}.}
	\label{fig:gzoltar}
\end{center}
\end{figure}

GZoltar\footnote{The GZoltar Project -- \url{http://www.gzoltar.com/}}~\cite{Riboira2010,Riboira2011a} is a framework for automating testing and debugging projects written in Java. It is an Eclipse-based, Java implementation
of Zoltar that integrates with frameworks such as JUnit. It also provides powerful hierarchical
visualization and interaction options to developers (such as a sunburst view, see Figure~\ref{fig:gzoltar}).

The GZoltar tool was the base of André Riboira's MSc thesis~\cite{Riboira2011} and is under active development at Faculdade de Engenharia da Universidade do Porto. The work detailed in this thesis is aimed at improving this tool. Further information about the GZoltar project is available in Section~\ref{sec:gzoltar_toolset}.

\section{Reasoning Approaches} 
\label{sec:reasoning_approaches}

Reasoning approaches to fault localization use prior knowledge of the system, such as required component behavior and interconnection, to build a model of the system behavior. An example of a reasoning technique is \ac{MBD} (see, e.g.,~\cite{deKleer87}).

\subsection{Model-Based Diagnosis} 
\label{ssec:model_based_diagnosis}

In \ac{MBD}, a diagnosis is obtained by logical inference from the {\em static} model of the system, combined with a set of run-time observations. Traditional \ac{MBD} systems require the model to be supplied by the system's designer, whereas the description of the observed behavior is gathered through direct measurements. The difference between the behavior described by the model and the observed behavior can then be used to identify components that may explain possible deviations from normal behavior~\cite{Mayer2007}.

In practice, as a formal description of the program is required, the task of using \ac{MBD} can be difficult. This is particularly due to (1) the large scope of current software projects, where formal models are rarely made available, and (2) the maintenance problems that arise throughout development, since changes in functionality are likely to happen. Furthermore, formal models usually do not describe a system's complete behavior, being restricted to a particular component of the system.


\subsection{Model-Based Software Debugging} 
\label{sub:model_based_software_debugging}

In order to address some of the issues that traditional \ac{MBD} has, \ac{MBSD} exchanges the roles of the model and the observations. In this technique, instead of requiring the designer to formally specify the intended behavior, a model is automatically inferred from the actual program. This means that the model reflects all the faults that exist in the program. The correct behavior specification in this technique is described in the system's test cases, which specify the expected output for a certain input.

Well-known approaches to \ac{MBSD} include the approaches of Friedrich, Stumptner, and Wotawa \cite{vhdl,mbd_vhdl}, Nica and Wotawa~\cite{wotawa08}, Wotawa, Stumptner, and Mayer~\cite{Wotawa2002}, and Mayer and Stumpter~\cite{Mayer2007}. However, \ac{MBSD} has problems concerning scalability -- the computational effort required to create a model of a large program forbids its use in real-life applications~\cite{Mayer2008}.



\section{Discussion} 
\label{sec:discussion}




Currently, the most effective debugging tools are the \acp{SDT}. These tools, with minimal effort from the user,
return a ranked list of potential faulty locations. By comparison, traditional debugging tools require more user 
effort for locating faults, and are fairly \emph{ad-hoc}.

\acp{SDT} not only require less effort, but they also help improve the debugging process by automating it. This 
is particularly useful for regression testing.

However, \acp{SDT} have some flaws regarding performance, as the \ac{SUT} has to be instrumented. Studies have
shown that instrumentation can hit execution time by as much as 50\%~\cite{Yang2006}, so this approach of fault
localization is particularly inefficient for large, real systems, that contain hundreds of thousands of \acp{LOC}. Also, while parallelization can help minimize the impact of instrumentation, we may not be able to use it in every situation (and particularly while dealing with resource constrained projects).

Other debugging techniques, use reasoning approaches to debugging, such as \ac{MBSD}. However, the computational
effort required to create a model of a large application is very high. 


\chapter{Dynamic Code Coverage} 
\label{cha:dcc}

In this chapter, we present a motivational example showing why traditional \ac{SFL} approaches can be inefficient, mainly due to the overhead caused by instrumenting every line of code. Afterwards, we propose and algorithm, coined \acf{DCC}, that mitigates those inefficiencies by gradually adjusting the instrumentation granularity of each software component of the \ac{SUT}.

\section{Motivational Example} 
\label{sec:motivational_example}

Suppose a program responsible for controlling a television set is being debugged. Consider that such program has three main high-level modules:

\begin{enumerate}
	\item Audio and video processing;
	\item Teletext decoding and navigation;
	\item Remote-control input.
\end{enumerate}

If one is to use \ac{SFL} to pinpoint the root cause of observed failures, hit spectra for the entire application have to be gathered. Furthermore, the hit spectra have to be of a fine granularity, such as \ac{LOC} level, so that the fault is more easily located. 

\begin{figure}
\begin{center}
	\includegraphics[width=1\textwidth]{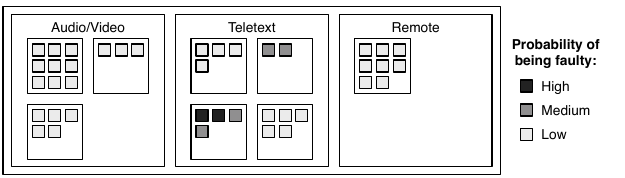}
	\caption{\ac{SFL} output example.}
	\label{fig:example-sota}
\end{center}
\end{figure}

An output of the \ac{SFL} technique applied to this specific example can be seen in Figure~\ref{fig:example-sota}. The smaller squares represent each \ac{LOC} of the program, which are grouped into methods, and then into the three main modules of the program under test.

As seen in Figure~\ref{fig:example-sota}, every \ac{LOC} in the program has an associated fault coefficient that represents the probability of that component being faulty. In this example, the bottom-left function of the teletext decoding and navigation module has two \acp{LOC} with high probability of being faulty, and other two with medium probability. The upper-right function of the teletext module also contains two medium probability \acp{LOC}. 

There are, however, many \acp{LOC} with low probability of containing a fault. In fact, in some methods, and even entire modules, such as the audio/video processing module and the remote-control module, all components have low probability of being at fault. Such low probability is an indication that the fault might be located elsewhere, and thus these components need not to be inspected first.  

As \ac{SFL} needs to have information about the entire program spectra to perform an analysis on the most probable fault locations, this can lead to scalability problems, as every \ac{LOC} has to be instrumented. As previously stated in Section~\ref{sub:code_coverage}, instrumentation can hit execution time by as much as 50\% in code coverage tools that use similar instrumentation techniques~\cite{Yang2006}. As such, fault localization that uses hit spectra is acceptable for debugging software applications, but may be impractical for large, real-world, and resource-constrained projects that contain hundreds of thousands of \acp{LOC}.

In order to make \ac{SFL} amenable to large, real, and resource-constrained applications, a way to avoid instrumenting the entire program must be devised, while still having a fine granularity for the most probable locations in the fault localization results.


\section{Dynamic Code Coverage Algorithm} 
\label{sec:dynamic_code_coverage}

In order to solve the scaling problems that automated fault localization tools have, it is proposed a \ac{DCC} approach. 
This method uses, at first, a coarser granularity level of instrumentation and progressively increases the instrumentation detail of potential faulty components.

\ac{DCC} is shown in Algorithm~\ref{alg:dcc-algorithm}. It takes as parameters $System$, $TestSuite$, $InitialGranularity$ and $FinalGranularity$. These parameters correspond to the \ac{SUT}, its test case set, and the initial and final instrumentation detail levels, respectively.

\begin{algorithm}[H]
\begin{algorithmic}[1]
\Procedure{DCC}{$System, TestSuite, InitialGranularity, FinalGranularity$}
\State $\mathcal{R} \gets \varnothing$
\State $\mathcal{F} \gets System$
\State $\mathcal{T} \gets TestSuite$
\State $\mathcal{G} \gets InitialGranularity$

\Repeat\label{alg:dcc-algorithm:loop-start}
	\State $\Call{Instrument}{\mathcal{F},\mathcal{G}}$
	\State $(A,e) \gets \Call{RunTests}{\mathcal{T}}$\label{alg:dcc-algorithm:run-tests}
	\State $\mathcal{C} \gets \Call{SFL}{A,e}$\label{alg:dcc-algorithm:sfl}
	\State $\mathcal{F} \gets \Call{Filter}{\mathcal{C}}$
	\State $\mathcal{R} \gets \Call{Update}{\mathcal{R},\mathcal{F}}$
	\State $\mathcal{T} \gets \Call{NextTests}{TestSuite,A,\mathcal{F}}$\label{alg:dcc-algorithm:next-tests}
	\State $\mathcal{G} \gets \Call{NextGranularity}{\mathcal{F}}$
\Until{ $\Call{IsFinalGranularity}{\mathcal{F}, FinalGranularity}$ }

\State \textbf{return} $\mathcal{R}$
\EndProcedure
\end{algorithmic}
\caption{\acl{DCC}.}
\label{alg:dcc-algorithm}
\end{algorithm}

First, an empty report $\mathcal{R}$ is created. After that, a list of the components to instrument $\mathcal{F}$ is initialized with all $System$ components. Similarly, the list of test cases to run in each iteration $\mathcal{T}$ is initialized with all test cases in $TestSuite$. An initial granularity $\mathcal{G}$ is also initialized with the desired initial exploration granularity $InitialGranularity$, which can be set from a class level to a \ac{LOC} level.

After the initial assignments, the algorithm will start its iteration phase in line~\ref{alg:dcc-algorithm:loop-start}. At the start of each iteration, every component in the list $\mathcal{F}$ is instrumented with the granularity $\mathcal{G}$ with the method $\Call{Instrument}{}$. What this method does is to alter these components so that their execution is registered in the program spectra.

Afterwards, the test cases $\mathcal{T}$ are run with the method $\Call{RunTests}{}$. Its output is a hit spectra matrix $A$ for all the previously instrumented components, and the error vector $e$, that states what tests passed and what tests failed. As explained in Section~\ref{sec:statistical_debugging}, these are the necessary inputs for spectrum-based fault localization, performed in line~\ref{alg:dcc-algorithm:sfl}. This $\Call{SFL}{}$ method calculates, for each instrumented component, its failure coefficient using the Ochiai coefficient, previously shown in equation~\ref{eq:ochiai}.

Following the fault localization step, the components are passed through a $\Call{Filter}{}$ that eliminates the low probability ones according to a set threshold, and the list $\mathcal{F}$ is updated, as well as the fault localization report $\mathcal{R}$. 

In line~\ref{alg:dcc-algorithm:next-tests}, the test case set is updated to run only the tests that touch the current components $\mathcal{F}$. Such tests can be retrieved by analyzing the coverage matrix $A$.

The last step in the iteration is to update the instrumentation granularity for next iterations. Method $\Call{NextGranularity}{}$ finds the coarser granularity in all the components of list $\mathcal{F}$, and updates that granularity to the next level of detail.

Every iteration is tested for recursion with $\Call{IsFinalGranularity}{}$, that returns true if every component in the list $\mathcal{F}$ is at the desired final granularity defined in $FinalGranularity$. This final granularity can be of different detail levels, such as method level or \ac{LOC} level, according to the needs of the software project being tested. If the $\Call{IsFinalGranularity}{}$ condition is not met, a new iteration is performed. 

Lastly, the \ac{DCC} algorithm returns the fault localization report $\mathcal{R}$. $\mathcal{R}$ contains diagnosis candidates of different granularity, typically with the top ones at the statement-level granularity.

\begin{figure}[h]
\begin{center}
	\includegraphics[width=0.5\textwidth]{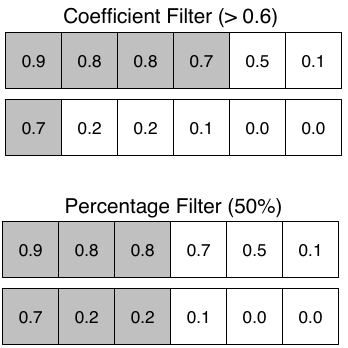}
	\caption{Component filters.}
	\label{fig:filters}
\end{center}
\end{figure}

\ac{DCC}'s performance is very dependent on the $\Call{Filter}{}$ function, which is responsible to decide whether or not it is required to zoom-in\footnote{In this context, zooming-in is to explore the inner components of a given component.} in a given component. Although many filters may be plugged into the algorithm, in this thesis we study the impact of two filters (see Figure~\ref{fig:filters} for an illustration):
\begin{itemize}
	\item{Coefficient filter $C_f$ -- components above the \ac{SFL} coefficient threshold $C_f$ are expanded.}
	\item{Percentage filter $P_f$ -- the first $P_f$\% components are expanded.}
\end{itemize}

\begin{figure}[h]
\centering
	\subfloat[First iteration] {
		\includegraphics[width=1\textwidth]{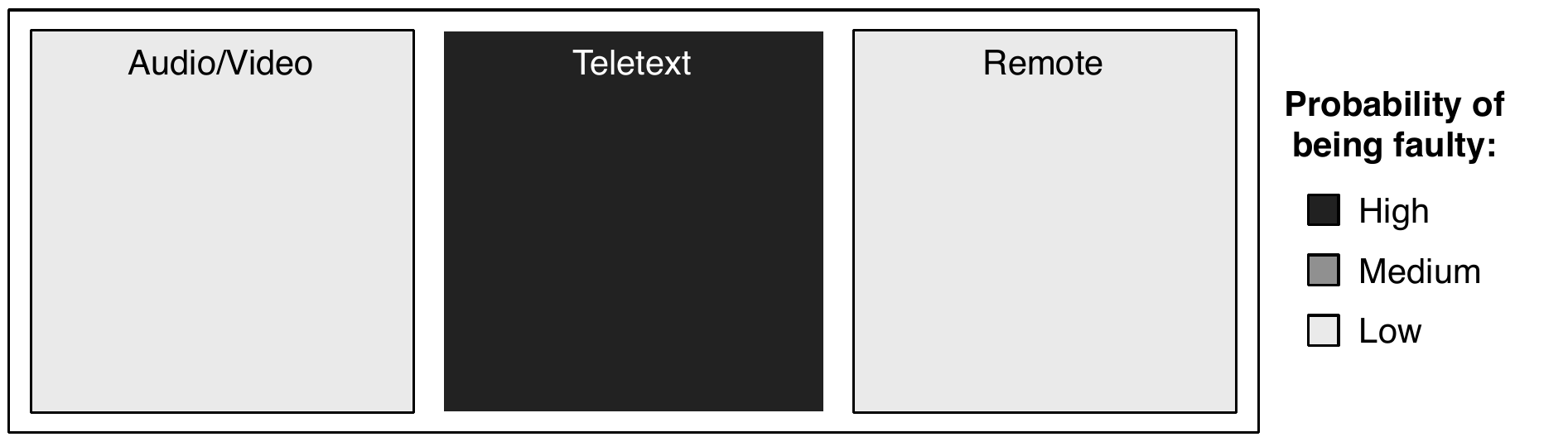}
		\label{fig:example-dcc1}
	}
	
	\subfloat[Second iteration] {
		\includegraphics[width=1\textwidth]{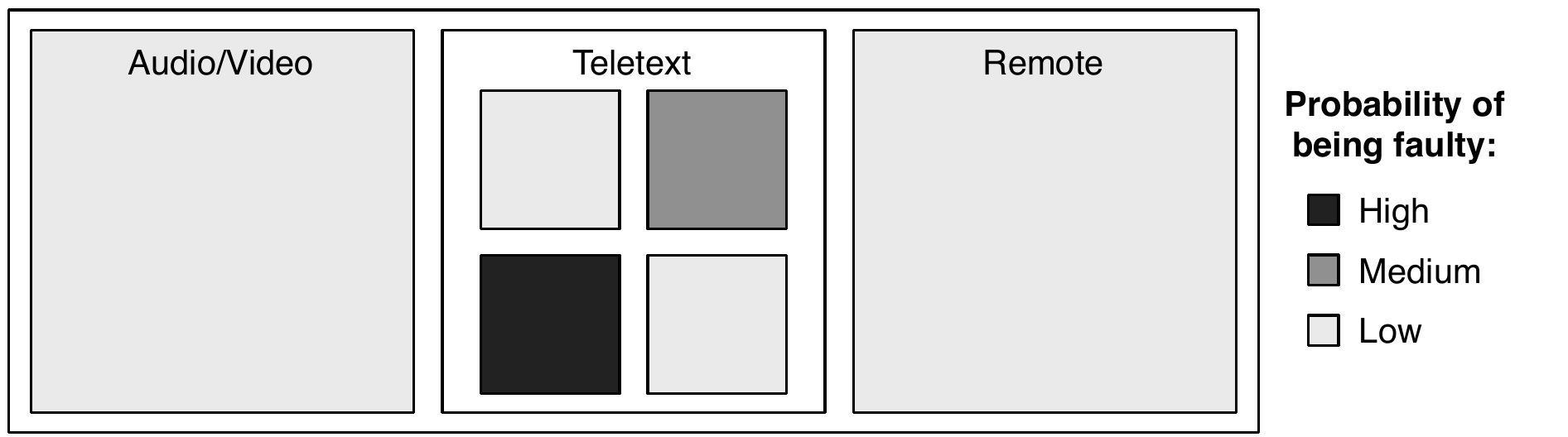}
		\label{fig:example-dcc2}
	}
	
	\subfloat[Third iteration] {
		\includegraphics[width=1\textwidth]{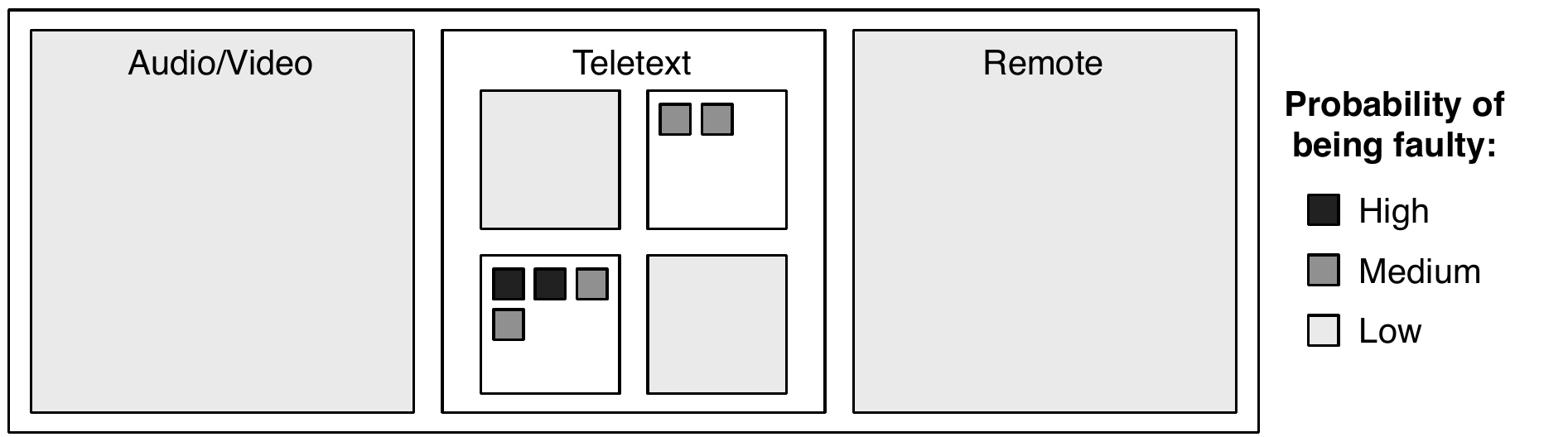}
		\label{fig:example-dcc3}
	}
	\caption{\ac{DCC} output example.}
	\label{fig:example-dcc}
\end{figure}

To illustrate the overhead reduction, let us revisit the motivational example given in Section~\ref{sec:motivational_example}. If use the \ac{DCC} approach to debug this program, we get the output of each iteration of the algorithm as shown in Figure~\ref{fig:example-dcc}. In this example, a filter responsible for not exploring components with low probability of containing faults is being used. In particular, the algorithm executes as follows:

\begin{enumerate}
	\item{The three modules -- Audio/Video, Teletext, and Remote -- are instrumented at the module level. Upon running the tests and \ac{SFL}, the only component with high probability of being faulty is the Teletext module. See Figure~\ref{fig:example-dcc1}.}
	\item{The Teletext module is instrumented at a method level. After that, the tests that touch the Teletext module are run. Fault localization states that the upper-right (UR) and the bottom-left (BL) functions have medium and high probability of being faulty, respectively. See Figure~\ref{fig:example-dcc2}.}
	\item{The UR and BL functions are instrumented at the \ac{LOC} level. After the tests that touch those functions are run and fault localization is performed, every \ac{LOC} in those functions has an associated fault coefficient. As all the non-filtered components are of \ac{LOC} granularity, the execution is terminated. See Figure~\ref{fig:example-dcc3}.}
\end{enumerate}

This approach, besides only reporting \acp{LOC} which are more likely to contain a fault, also needs to instrument less software components -- 13 in total. Compared to the pure \ac{SFL} approach of Section~\ref{sec:motivational_example}, where 40 components were instrumented, \ac{DCC} has reduced instrumentation (thus, its overhead) by 67.5\%, while producing the same good results.


\section{Discussion} 
\label{sec:discussion}

\ac{DCC} is a fault localization algorithm that instruments software components at a low level of detail, and progressively increases the instrumentation detail of software components likely to be at fault.

The main advantages of our \ac{DCC} algorithm are twofold. 

\begin{enumerate}
	\item The instrumentation overhead in the program execution decreases. This is due to the fact that not every \ac{LOC} is instrumented -- only the \acp{LOC} most likely to contain a fault will be instrumented at that level of detail.
	\item In every iteration, the generated program spectra matrices, seen in line~\ref{alg:dcc-algorithm:run-tests} of Algorithm~\ref{alg:dcc-algorithm}, will be shorter in size when compared to traditional \ac{SFL}. That way, the fault coefficient calculation, described in Section~\ref{sec:statistical_debugging}, will be inherently faster, as there are fewer components to calculate.
\end{enumerate}

The iterative nature of the \ac{DCC} algorithm also provides some benefits. In each iteration, the algorithm is walking towards a solution, narrowing down the list of components which are likely to contain a fault. As such some information about those components can be made available, directing the developer to the fault location even before the algorithm is finished. Also, as low probability components are being filtered, the final report will be shorter, providing the developer with a more concise fault localization report.

The ability that \ac{DCC} has to stop at any desired granularity level of detail can also provide benefits. For instance, some defects may be successfully diagnosed upon inspecting the list of faulty methods in a project, so an instrumentation at a \ac{LOC} may not be always necessary. A more important use case of this cutoff ability, though, is to combine complementary fault localization methods. For example, \ac{DCC} can be used to obtain the top ranked methods in a program, and then employ the \ac{MBSD} technique, detailed in Section~\ref{sub:model_based_software_debugging}, only on these software components. This way, we are obtaining more accurate fault hypotheses due to the use of \ac{MBSD}, with significantly less computation required.

However, this algorithm still poses a couple of shortcomings. As \ac{DCC} relies on program hierarchy to devise different detail levels of instrumentation, some projects may not be suited to be debugged with this technique. In fact, while object-oriented programming languages, which have at least three clearly defined detail levels (\emph{i.e.}, class, method and \ac{LOC}), perform well with \ac{DCC}, other programming languages with different paradigms may not produce the same results in terms of instrumentation overhead reduction. 

An additional shortcoming is that, for smaller projects, \ac{DCC} may produce worse time execution results when compared with \ac{SFL}. This is due to the fact that these projects tend to produce denser program spectra matrices, because each test covers a considerable portion of the source code. Thus, the filtering operations will be ineffective as many components will have similar fault coefficients and the overhead of re-executing the majority of a project's test suite will be greater than instrumenting the project at a high level of detail.   


\chapter{Tooling} 
\label{cha:tooling}

In this chapter, the chosen tool to host the \ac{DCC} prototype -- GZoltar -- will be presented. Afterwards, some modifications and improvements made to the GZoltar tool will be detailed, as well as some of the most relevant implementation details of the \ac{DCC} prototype.

\section{GZoltar Toolset} 
\label{sec:gzoltar_toolset}

GZoltar\footnote{The GZoltar toolset can be found online at \url{http://gzoltar.com/}}, also mentioned in Section~\ref{sub:gzoltar}, is an Eclipse plugin that performs fault localization tasks using state-of-the-art fault localization algorithms. 

\begin{figure}[H]
\begin{center}
	\includegraphics[width=1\textwidth]{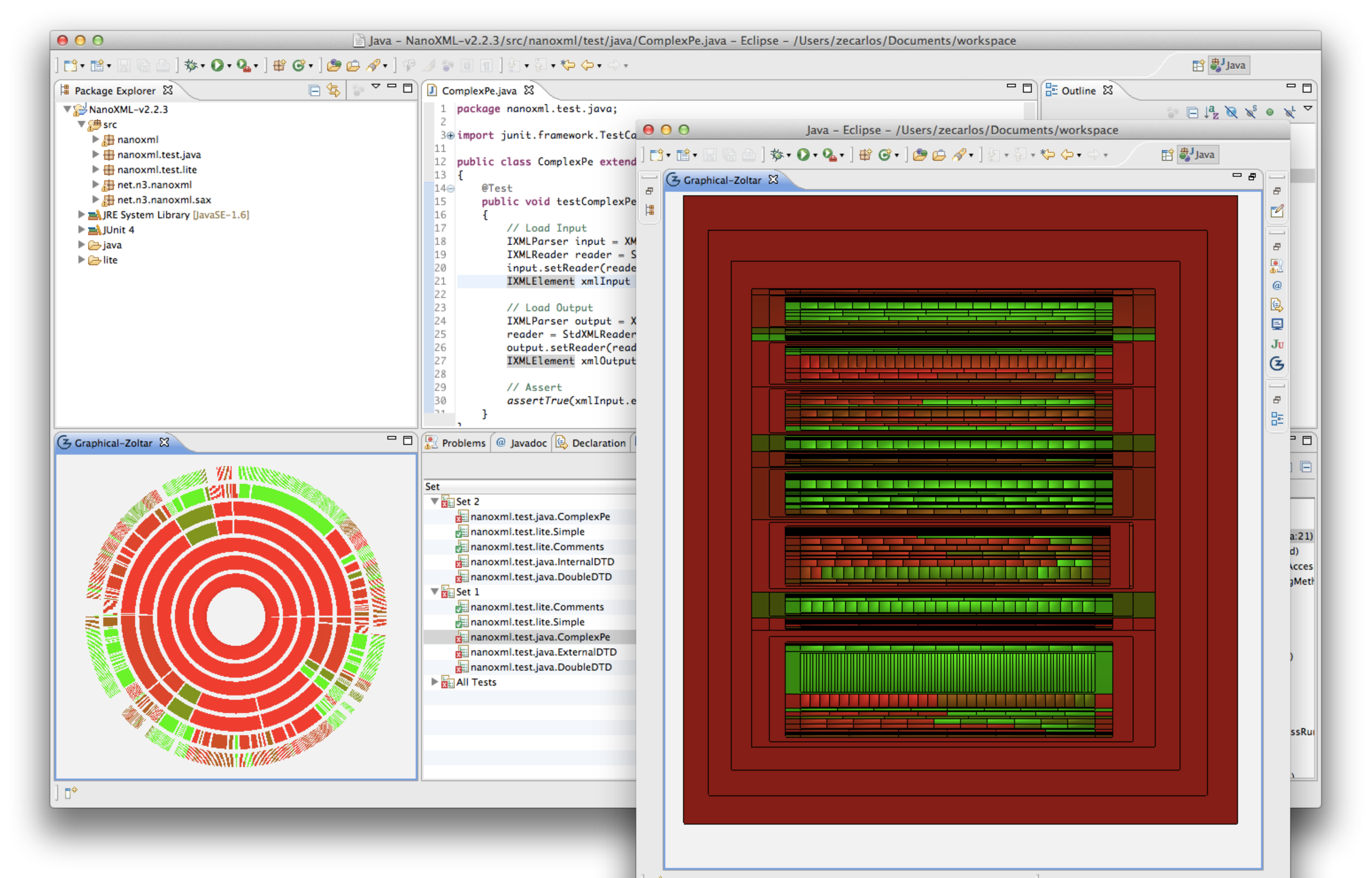}
	\caption{GZoltar's visualizations: Sunburst and Treemap.}
	\label{fig:gzoltar_visualizations}
\end{center}
\end{figure}

The GZoltar toolset implements \ac{SFL} with the Ochiai similarity coefficient (see Equation~\ref{eq:ochiai}), one of the best coefficients for this purpose. GZoltar also creates powerful and navigable diagnostic report visualizations, such as Treemap and Sunburst~\cite{Riboira2011} (see Figure~\ref{fig:gzoltar_visualizations}).

GZoltar is aimed at testing Java projects that use JUnit as their testing framework. Being an Eclipse (one of the most popular \acp{IDE}) plugin means that GZoltar can use many of the \ac{IDE}'s functionalities, such as detection of open projects in the workspace and improved interaction between the diagnostic report visualizations and the code editor (\emph{e.g.}, when a user clicks a certain line of code in the visualization, the corresponding file is opened in the editor, and the cursor is positioned in the desired line).

\begin{figure}[H]
\begin{center}
	\includegraphics[width=1\textwidth]{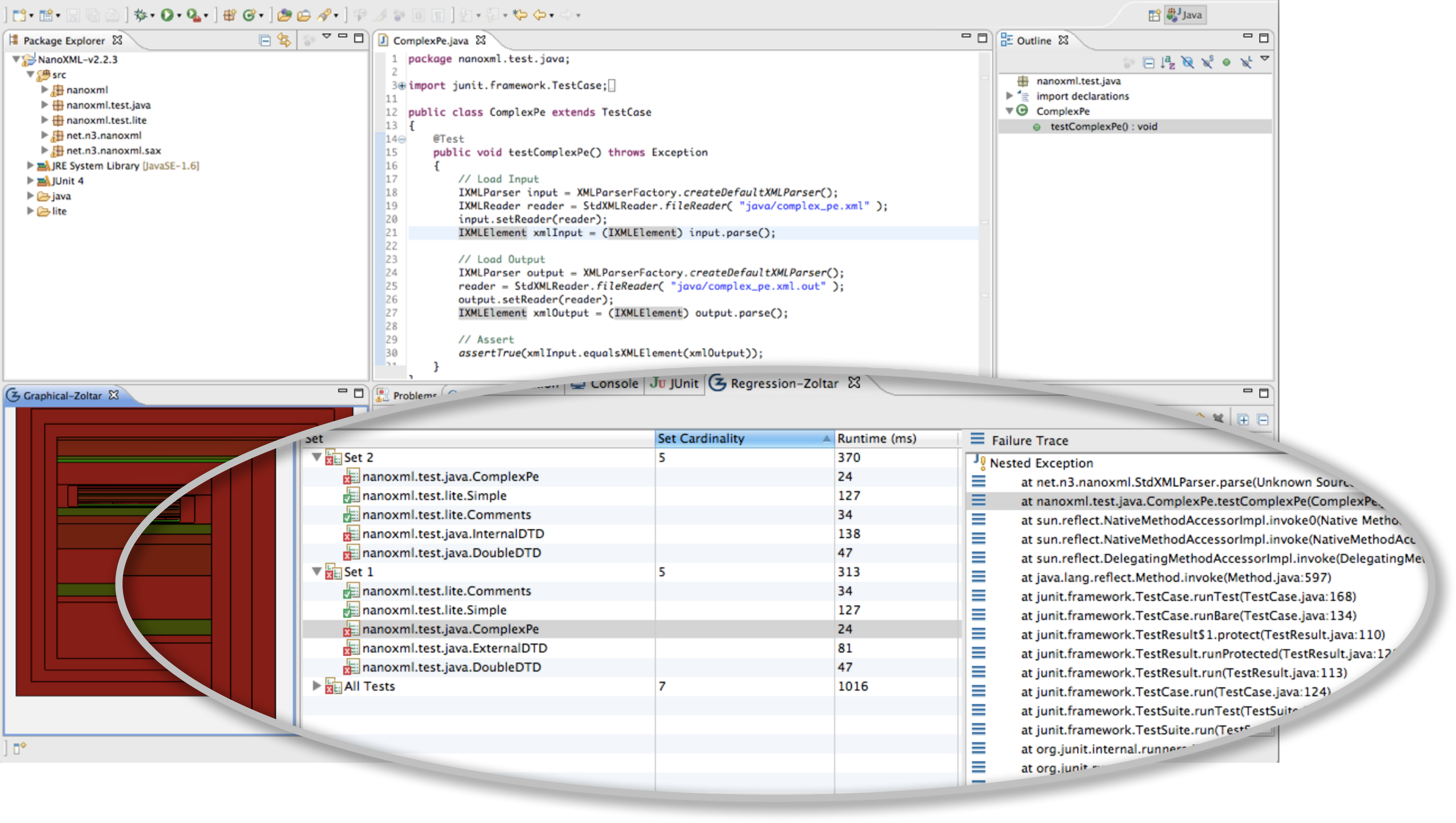}
	\caption{RZoltar's interface.}
	\label{fig:rzoltar_interface}
\end{center}
\end{figure}

Besides fault localization, GZoltar also provides a test suit reduction and prioritization tool, coined RZoltar (see Figure~\ref{fig:rzoltar_interface}). This tool minimizes the size of the original test suite using constraint-based approaches~\cite{STVR:STVR430}, while still guaranteeing the same code coverage. Also, RZoltar allows the user to prioritize the minimized test suites by cardinality or by execution time. 

\section{Modifications and Improvements} 
\label{sec:modifications_and_improvements}

In order to implement the \ac{DCC} prototype on top of GZoltar, some improvements and modifications had to be made to the underlying tool.

The first considerable change in GZoltar's architecture was the test case execution. Originally, test cases were run in the same \ac{JVM} that hosted Eclipse and all its plugins. This has several disadvantages. Firstly, every class in the project had to be explicitly loaded before running any test, because the classes were not visible in the \ac{JVM}'s \emph{classpath}. Secondly, since the \ac{JVM} was already running, some parameters could not be customized (\emph{e.g.}, the maximum heap size that can be allocated). Lastly, as the \ac{JVM}'s working directory is not the same as the system being tested, file operations with relative path names would not work correctly.

These issues were resolved by creating an external process that would spawn a new \ac{JVM} in the \ac{SUT}'s working directory, with its corresponding \emph{classpath}. After running the test cases, the test results and coverage are sent to GZoltar via a socketed connection.

Another issue with the original version of GZoltar is the way how the code instrumentation is handled. GZoltar used the code coverage library JaCoCo\footnote{JaCoCo -- \url{http://www.eclemma.org/jacoco/index.html}} to obtain the coverage traces (\emph{i.e.}, hit spectra) needed to perform the fault localization analysis. However, JaCoCo uses a \ac{LOC} level of instrumentation, and is unable to gather coverage information without this fine-grained level of detail. To prevent this, we discarded JaCoCo and created a coverage tool able to gather traces from three different granularities: (1) class level, (2) method level and (3) \ac{LOC} level.

This coverage tool makes use of the \emph{java.lang.instrument} framework that was incorporated into the version 1.5 of the \ac{JVM}. This framework allows the user to attach an agent that is able to intercept and modify a class' \emph{bytecode} before it is loaded by the Java \emph{ClassLoader}. This way, the instrumentation code, written with the aid of ASM\footnote{ASM -- \url{http://asm.ow2.org/}}, a Java \emph{bytecode} manipulation framework, can be inserted at each class load time.

\begin{figure}[h]
\begin{center}
	\includegraphics[width=0.35\textwidth]{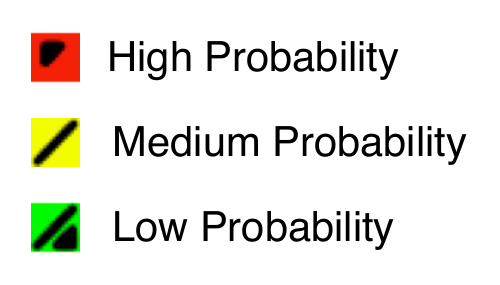}
	\caption{Statement failure probability markers.}
	\label{fig:probability_markers}
\end{center}
\end{figure}

Finally, as an accessibility improvement, GZoltar can now generate a list of markers that are placed on the code editor's vertical ruler, indicating the respective line of code's probability of being faulty when hovering the mouse. These markers, as we can see in Figure~\ref{fig:probability_markers}, can be of three different types: 
(1) red for the top third statements most likely to contain a fault, 
(2) yellow for the middle third statements, and 
(3) green for the bottom third statements. 
Every marker also has an embedded ColorADD\footnote{ColorADD color identification system -- \url{http://coloradd.net/}} symbol, in order to help colorblind people distinguish between markers.
These annotation markers are also displayed on the Eclipse ``Problems'' view.


\section{Dynamic Code Coverage Prototype} 
\label{sec:dynamic_code_coverage_prototype}

In this section, some of the most relevant implementation details of the \ac{DCC} prototype are presented.

The instrumentation for coverage gathering for each granularity level works as follows:

\begin{itemize}
	\item Class level -- an \emph{INVOKESTATIC} instruction is inserted at the beginning of the class initialization method (also known as \emph{<clinit>}), every constructor method, and every public static method. The inserted instruction will call a publicly available method called \emph{logClass} with the class details as parameters (\emph{i.e.}, name and package). This method will register that the class was touched by the execution.
	\item Method level -- every method of a certain class will be inserted with an \emph{INVOKESTATIC} instruction that calls \emph{logMethod}. This function takes as parameters the method information, namely its class, package and signature, and registers that the corresponding method was executed.
	\item \ac{LOC} level -- similarly to the granularities detailed previously, an invoke instruction is also inserted. In this granularity level, the instrumented instructions are placed at the beginning of each line of code, and call \emph{logLine} (that registers a hit whenever the line is executed). One thing to note is, similarly to other code coverage tools, line coverage information can only be gathered if the classes are compiled with debug information enabled, so that source line tables that map each line to their corresponding \emph{bytecode} instructions are available.
\end{itemize}

Some constructs, regardless of the set granularity level, are ignored while performing the instrumentation. Such is the case of synthetic constructs, which are introduced by the compiler and do not have a corresponding construct in the source code~\cite[p. 338]{jlspec}. Synthetic methods are generated for various purposes, e.g., to create bridge methods to ensure that type erasures and covariant return types are handled correctly.

Another key implementation detail worth mentioning is how the component filtering is handled. Filtering is one of the most important steps to ensure that the \ac{DCC} algorithm performs well. Because \ac{DCC} is a new concept, new filters should be easy to create and to replace, so that we can analyze their impact and quickly make adjustments in order to fine tune the algorithm's performance. As such, filters use a strategy pattern~\cite[pp. 315-323]{GoF}, and can easily be interchanged. 


\section{Discussion} 
\label{sec:discussion}

Some improvements and modifications were made to the GZoltar toolset, namely to provide more control over the testing and instrumentation tasks. It is worth to note that all improvements presented in Section~\ref{sec:modifications_and_improvements} have already been deployed to the development branch of GZoltar. In fact, the newest version of the tool (version 3.2.0, as of this writing) already contains all these modifications.

With these improvements, the creation of a \ac{DCC} prototype was made possible. In the next chapter, the validity of the \ac{DCC} approach will be evaluated using the implementation detailed in the previous sections.



\chapter{Empirical Evaluation} 
\label{cha:empirical_evaluation}

In this chapter, we evaluate the validity and performance of the \ac{DCC} approach for real projects. First, we introduce the programs under analysis and the evaluation metrics. Then, we discuss the empirical results and finish this section with a threats to validity discussion.

\section{Experimental Setup} 
\label{sec:experimental_setup}

For our empirical study, four subjects written in Java were considered:
\begin{itemize}
	\item{\texttt{NanoXML}\footnote{NanoXML -- \url{http://devkix.com/nanoxml.php}} -- a small XML parser.}
	\item{\texttt{org.jacoco.report} -- report generation module for the \texttt{JaCoCo}\footnote{JaCoCo -- \url{http://www.eclemma.org/jacoco/index.html}} code coverage library.}
	\item{\texttt{XML-Security} -- a component library implementing XML signature and encryption standards. This library is part of the Apache Santuario\footnote{Apache Santuario -- \url{http://santuario.apache.org/}} project.}
	\item{\texttt{JMeter}\footnote{JMeter -- \url{http://jmeter.apache.org/}} -- a desktop application designed to load test functional behavior and measure performance of web applications.}
\end{itemize}

The project details of each subject are in Table~\ref{tab:experimental-subjects}. The \ac{LOC} count information was gathered using the metrics calculation and dependency analyzer plugin for Eclipse \texttt{Metrics}\footnote{Metrics -- \url{http://metrics.sourceforge.net/}}. Test count and coverage percentage were collected with the Java code coverage plugin for Eclipse \texttt{EclEmma}\footnote{EclEmma -- \url{http://www.eclemma.org/}}.

\begin{table}[h]
	\centering

\begin{tabular}{|c|c|c|c|c|}
\hline
\textbf{Subject} & \textbf{Version} & \textbf{\acp{LOC} ($M$)} & \textbf{Test Cases} & \textbf{Coverage}\\
\hline
\texttt{NanoXML} & 2.2.6 & 5393 & 8 & 53.2\% \\
\hline
\texttt{org.jacoco.report} & 0.5.5 & 5979 & 225 & 97.2\% \\
\hline
\texttt{XML-Security} & 1.5.0 & 60946 & 461 & 59.8\% \\
\hline
\texttt{JMeter} & 2.6 & 127359 & 593 & 34.2\% \\
\hline
\end{tabular}
	\caption{Experimental Subjects.}
	\label{tab:experimental-subjects}
\end{table}

To assess the efficiency and effectiveness of \ac{DCC} the following experiments were performed, using fifteen faulty versions per subject program. We injected one fault in each of the 15 versions:
\begin{itemize}
	\item \ac{SFL} without \ac{DCC}. This is the reference baseline.
	\item \ac{DCC} with constant value coefficient filters from 0 to 0.95, with intervals of 0.05.
	\item \ac{DCC} with percentage filters from 100\% to 5\%, with intervals of 5\%.
\end{itemize}

The metrics gathered were execution time, the size of the fault localization report, and the average \acp{LOC} needed to be inspected until the fault is located. The latter metric can be calculated by sorting the fault localization report by the value of the coefficient, and finding the injected fault's position. In this metric, we are assuming that the developer performs the inspection in an ordered manner, starting from the highest fault coefficient \acp{LOC}. 

As spectrum-based fault localization creates a ranking of components in order of likelihood to be at fault, we can retrieve how many components we still need  to inspect until we hit the faulty one. Let $d\in\{1,\ldots,K\}$, where $K$ is the number of ranked components and $K \le M$, be the index of the statement that we know to contain the fault. For all $j\in\{1,\ldots,M\}$, let $s_j$. Then the ranking position of the faulty statement is given by

\begin{equation}
\tau = \frac{|\{ j | s_j > s_d \}| + |\{ j | s_j \ge s_d \}| - 1}{2}
\end{equation}

\noindent
$|\{ j | s_j > s_d \}|$ counts the number of components that outrank the faulty one, and $|\{ j | s_j \ge s_d \}|$ counts the number of components that rank with the same probability as the faulty one plus the ones that outrank it.

We define quality of diagnosis as the effectiveness to pinpoint the faulty component. As said before, this metric represents the percentage of components that need not be considered when searching for the fault by traversing the ranking. It is defined as

\begin{equation}\label{eq:qd}
(1 - \frac{\tau}{K_{SFL}}) \cdot 100\%
\end{equation}

\noindent where $K_{SFL}$ is the number of ranked components of \ac{SFL} without \ac{DCC} -- the reference baseline.

The experiments were run on a 2.7 GHz Intel Core i7 MacBook Pro with 4 GB of RAM, running OSX Lion.


\section{Experimental Results} 
\label{sec:experimental_results}

\begin{figure}[t]
\centering
	\subfloat[Coefficient filter] {
		\includegraphics[width=0.7\textwidth]{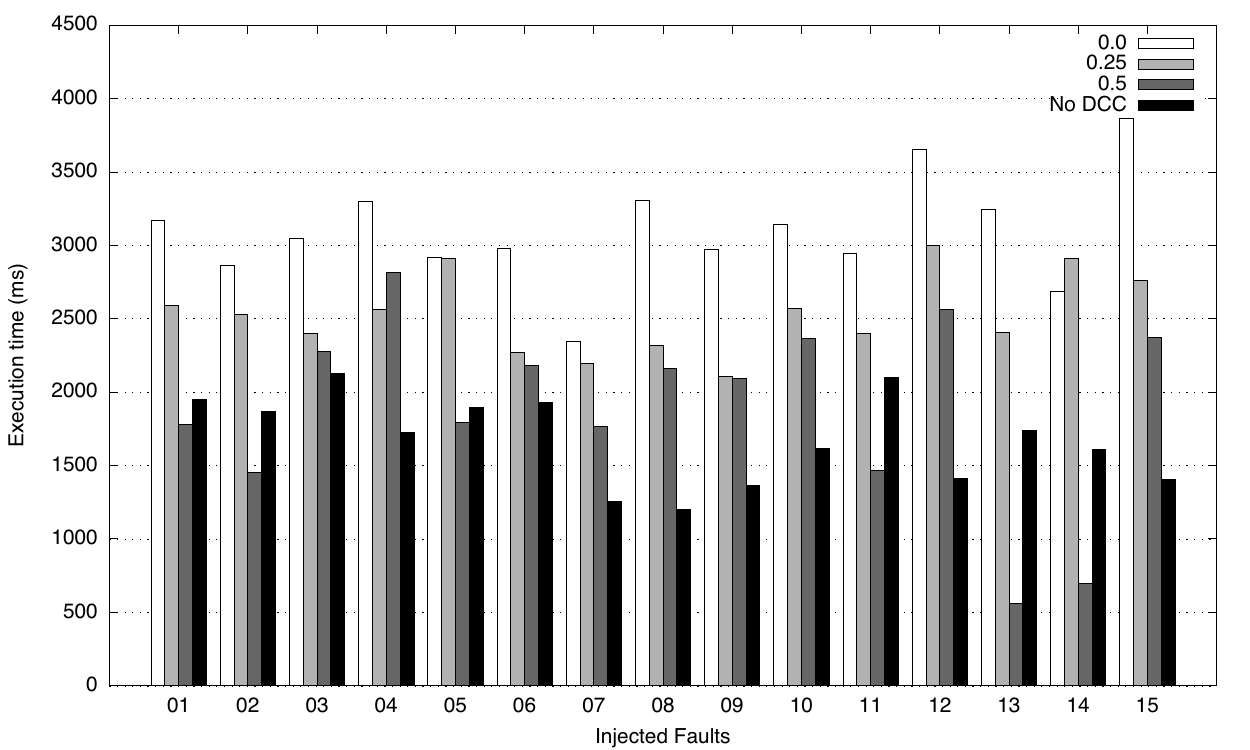}
		\label{fig:nanoxml-results-constantvalue}
	}

	\subfloat[Percentage filter] {
		\includegraphics[width=0.7\textwidth]{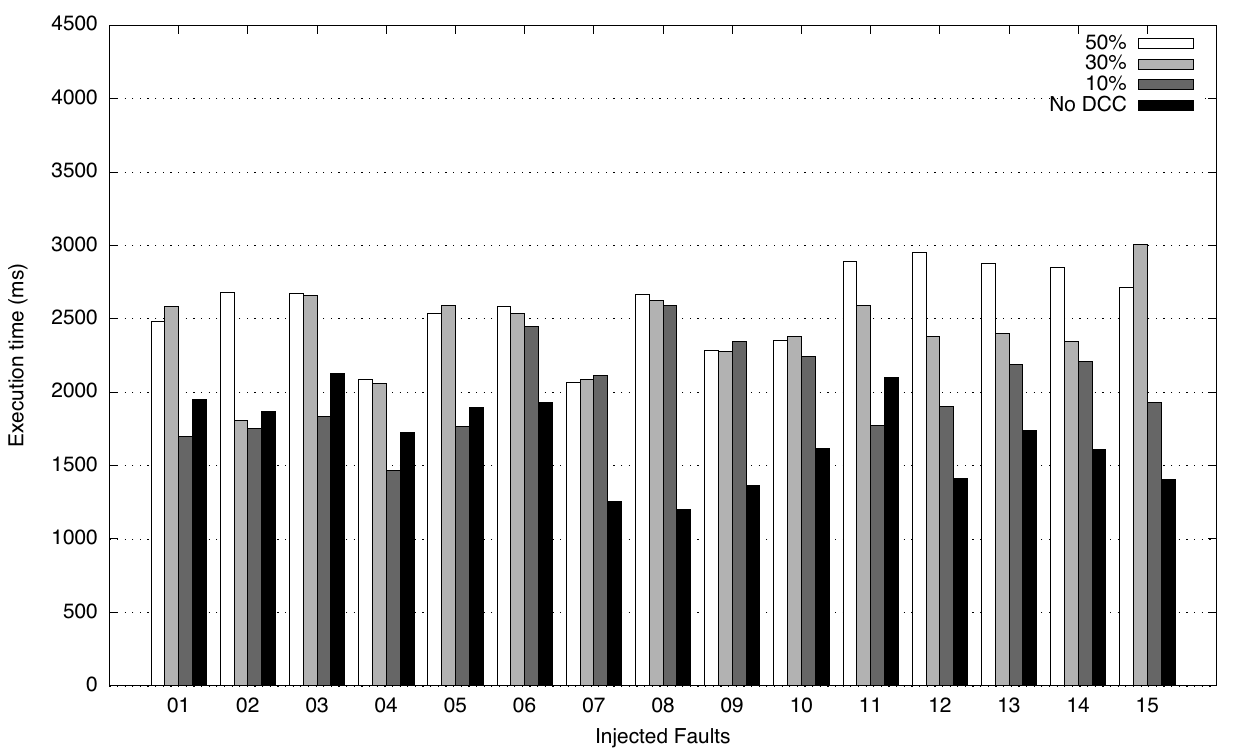}
		\label{fig:nanoxml-results-percentage}
	}
	\caption{\texttt{NanoXML} time execution results.}
	\label{fig:nanoxml-results}
\end{figure}

\begin{figure}[t]
\centering
	\subfloat[Coefficient filter] {
		\includegraphics[width=0.7\textwidth]{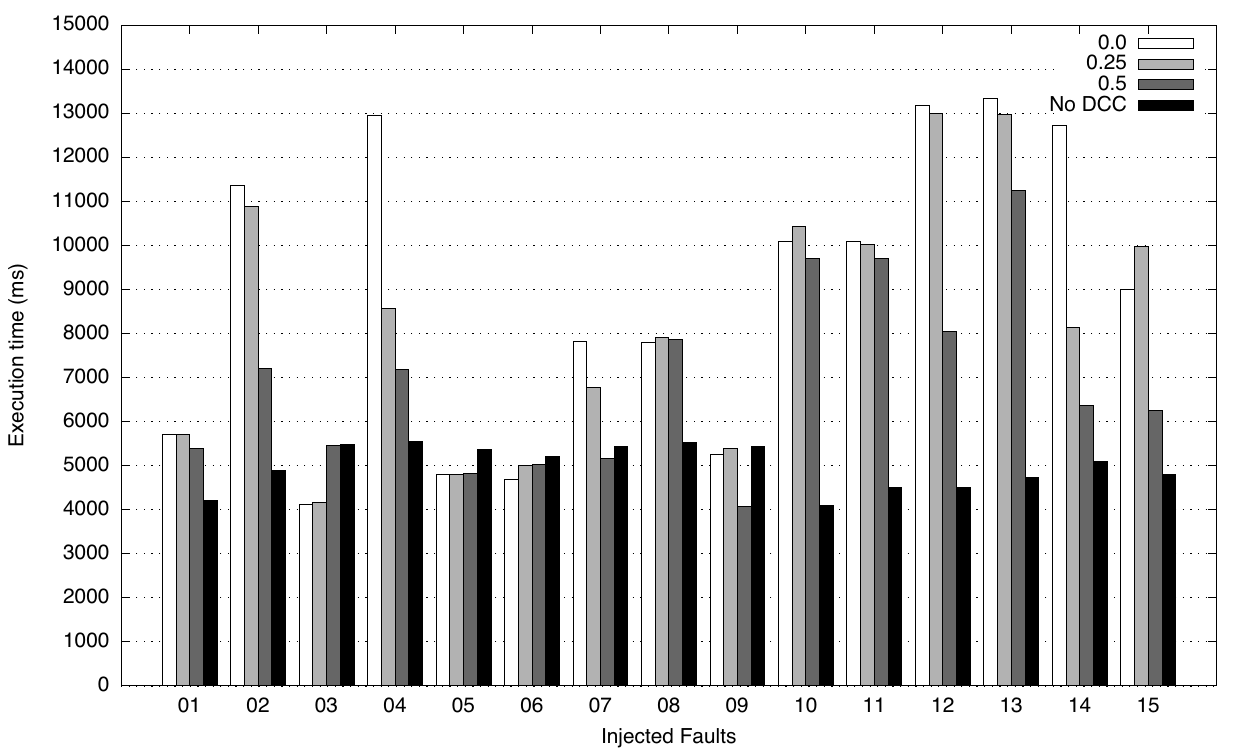}
		\label{fig:jacoco-results-constantvalue}
	}
	
	\subfloat[Percentage filter] {
		\includegraphics[width=0.7\textwidth]{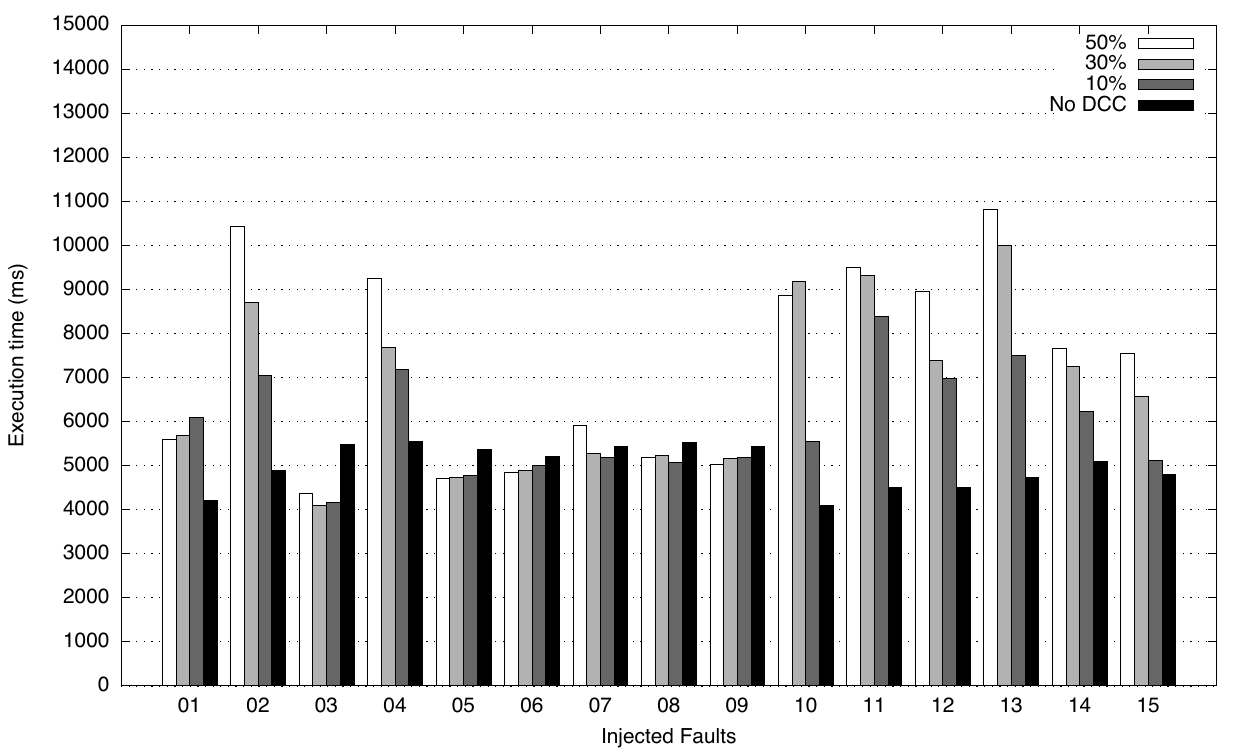}
		\label{fig:jacoco-results-percentage}
	}
	\caption{\texttt{org.jacoco.report} time execution results.}
	\label{fig:jacoco-results}
\end{figure}

\begin{figure}[t]
\centering
	\subfloat[Coefficient filter]{
		\includegraphics[width=0.7\textwidth]{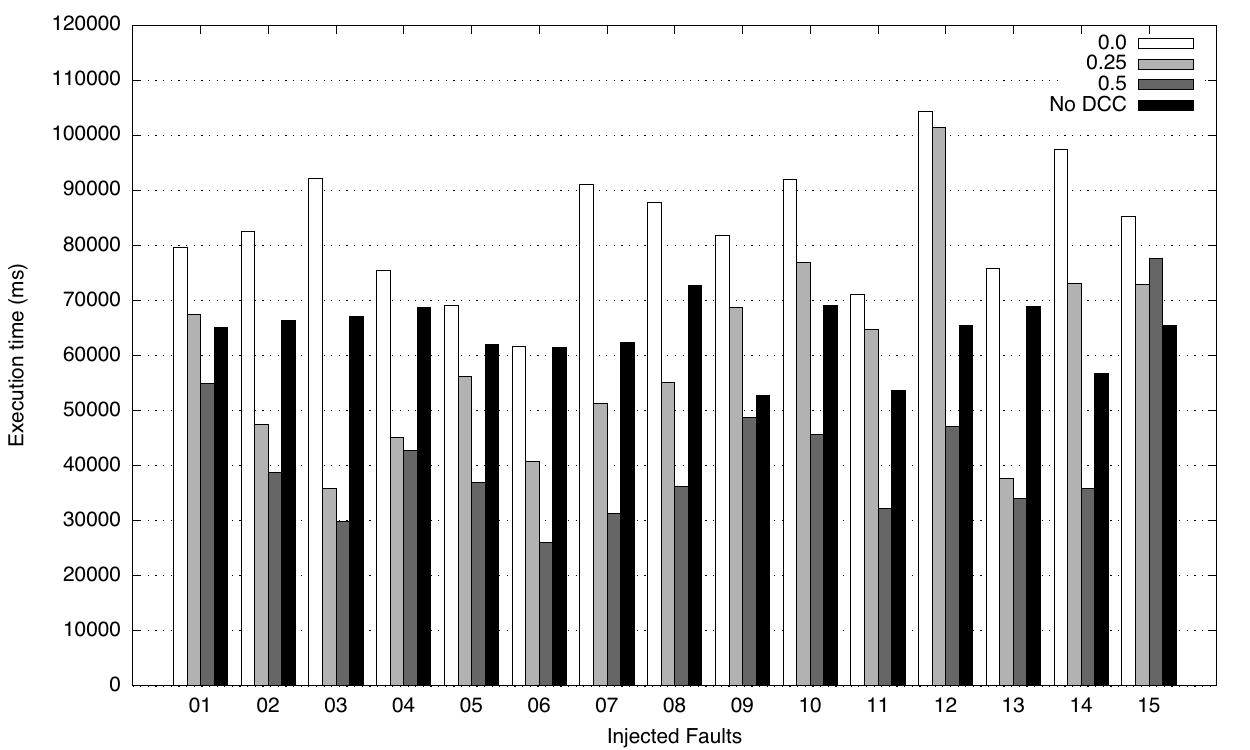}
		\label{fig:xmlsecurity-results-constantvalue}
	}
	
	\subfloat[Percentage filter]{
		\includegraphics[width=0.7\textwidth]{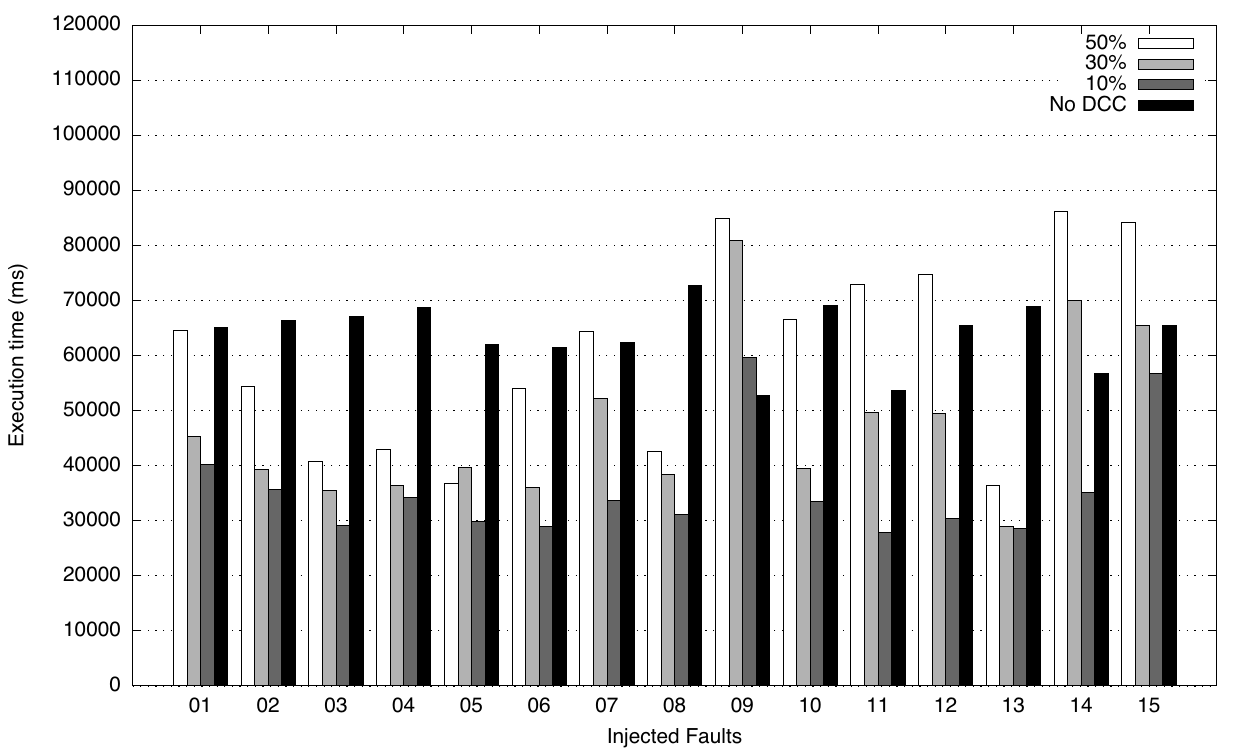}
		\label{fig:xmlsecurity-results-percentage}
	}
	\caption{\texttt{XML-Security} time execution results.}
	\label{fig:xmlsecurity-results}
\end{figure}

\begin{figure}[t]
\centering
	\subfloat[Coefficient filter]{
		\includegraphics[width=0.7\textwidth]{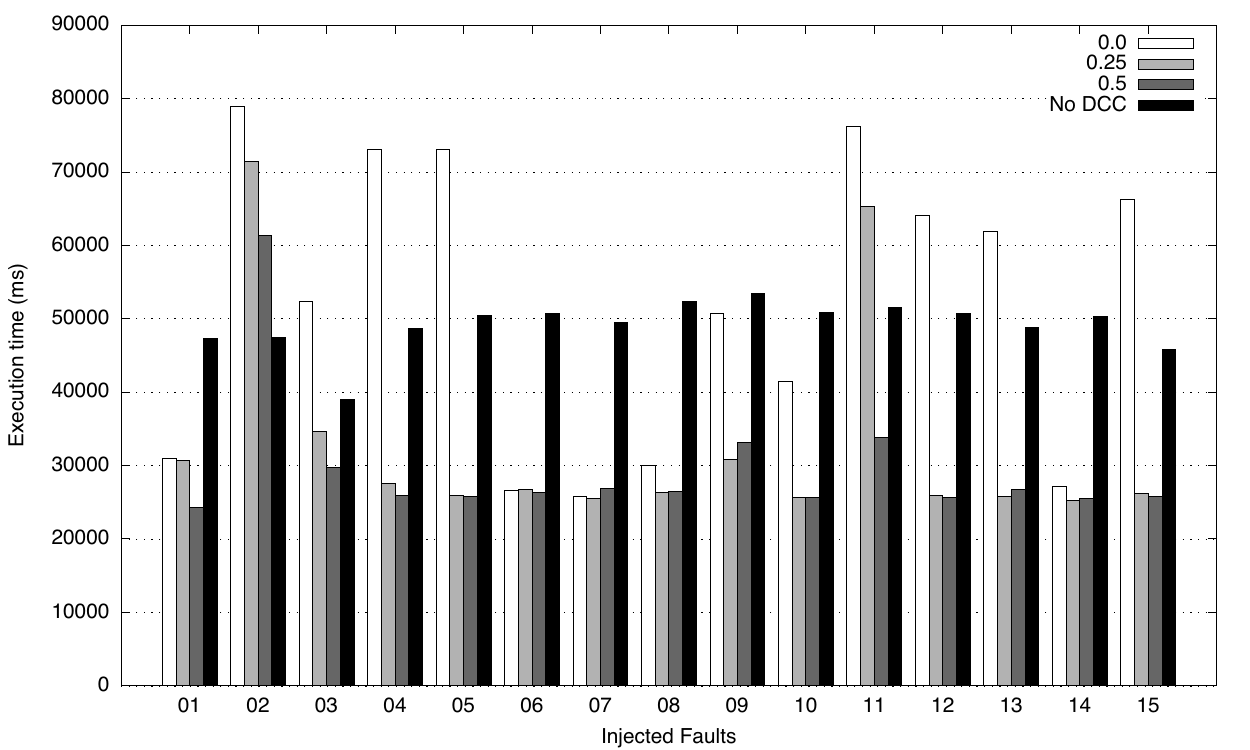}
		\label{fig:jmeter-results-constantvalue}
	}
	
	\subfloat[Percentage filter]{
		\includegraphics[width=0.7\textwidth]{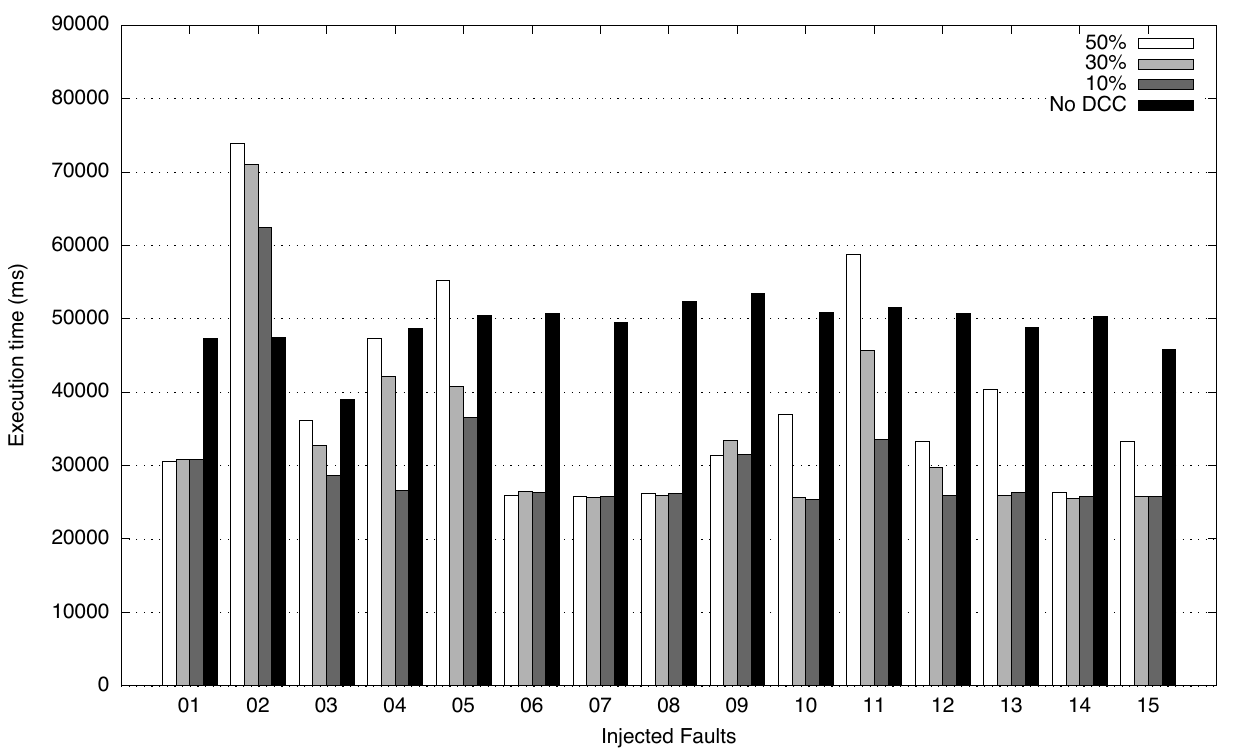}
		\label{fig:jmeter-results-percentage}
	}
	\caption{\texttt{JMeter} time execution results.}
	\label{fig:jmeter-results}
\end{figure}

Figures~\ref{fig:nanoxml-results},~\ref{fig:jacoco-results},~\ref{fig:xmlsecurity-results} and~\ref{fig:jmeter-results} summarize the overall execution time outcomes for all the experimental subjects. 

Each figure contains two plots, detailing the fault localization execution of each injected fault with \ac{DCC} using constant coefficient value filters and with \ac{DCC} using percentage filters, respectively. These filtering methods were previously detailed in Section~\ref{sec:dynamic_code_coverage}.

Due to space constraints, only three thresholds are shown for both filters: 0.0, 0.25 and 0.5 thresholds for the constant coefficient value filters ($C_f$) and 50\%, 30\% and 10\% thresholds for the percentage filters ($P_f$). To obtain a better understanding of the performance of each experiment, we also added, for every injected fault, the fault localization execution time of the \ac{SFL} without \ac{DCC} approach, labeled ``No DCC'' in the aforementioned figures. This way, \ac{DCC} approaches can be easily compared with the \ac{SFL} approach. Unless stated otherwise, every fault localization execution is able to find the injected fault (\emph{i.e.}, the resulting report contains the injected fault).

The first experimental subject to be analyzed was the \texttt{NanoXML} project, whose experiment results can be seen in Figure~\ref{fig:nanoxml-results}. Note that experiments 01, 11 and 13 for $C_f = 0.5$ (see Figure~\ref{fig:nanoxml-results-constantvalue}) and the experiments 01, 04, 11, 12, 13, 14, 15 for $P_f = 10\%$ (see Figure~\ref{fig:nanoxml-results-percentage}) were not able to find the injected faults.

As we can see from the experiment results, the \ac{DCC} approach underperforms the current \ac{SFL} method based in the execution time. Such results can be explained if we analyze the \texttt{NanoXML} project information in Table~\ref{tab:experimental-subjects}. This project, not only is rather small in size, but also has very few test cases. At the same time, it has a coverage of over 50\%. What this means is that some test cases, if not all, touch many different statements. As such, the generated program spectra matrices, detailed in Section~\ref{sec:statistical_debugging} will be rather dense. Because of this, many components will have similar coefficients, rendering the filtering operation ineffective: either discarding many different components, or keeping a lot of components to be re-instrumented and re-tested. 

The next analyzed subject was \texttt{org.jacoco.report}, part of the \texttt{JaCoCo} project. The filters $C_f = 0.5$ (see Figure~\ref{fig:jacoco-results-constantvalue}) and $P_f = 10\%$ (see Figure~\ref{fig:jacoco-results-percentage}) were both not able to find the injected faults in experiments 09 and 15. Also, the injected fault in experiment 02 was not found in $P_f = 10\%$.

This subject, despite having many more test cases than the previous project, still has some performance drops in some of the experiments. Upon investigating the fault localization reports of the lower performance experiments, we realized that their length can be as high as 950 statements in some experiments. This means that the set of test cases that touch the injected faulty statements can cover roughly 15\% of the entire project. Because of this, the same thing as the previous project happens: many components will have similar coefficients, rendering the expansions ineffective.

The following subject was the \texttt{XML-Security} project. Injected faults in experiments 03 and 08 were not found by $C_f = 0.5$ (see Figure~\ref{fig:xmlsecurity-results-constantvalue}). Experiment 08 also did not have its injected fault in $P_f = 10\%$ (see Figure~\ref{fig:xmlsecurity-results-percentage}).

The last subject was the \texttt{JMeter} project. Injected faults were not found by $C_f = 0.5$ (see Figure~\ref{fig:jmeter-results-constantvalue}) in experiments 01 and 11. Every fault was found with the percentage filters.

Both \texttt{XML-Security} and \texttt{JMeter} have better results when utilizing \ac{DCC}. There are mainly two reasons for these results. The first is the fact that the program spectra matrix is sparser. The other important reason is, as programs grow in size, the overhead of a fine-grained instrumentation, used in methodologies such as \ac{SFL}, is much more noticeable. In this kind of sizable projects (see project informations in Table~\ref{tab:experimental-subjects}), and if the matrix is sparse enough, it is preferable to re-run some of the tests, than to instrument every \ac{LOC} at the start of the fault localization process.  

These time execution results confirm our assumptions that \ac{DCC} can over-perform \ac{SFL} for larger projects, where the instrumentation overhead is heavily noticeable. In contrast, for smaller projects, \ac{DCC} does suffer in performance, mainly due to the fact that the overhead of re-running tests produces a bigger performance hit than the instrumentation granularity overhead. In fact, if we take into account all experiments for all four projects, there actually is an increase of execution time of 8\% ($\sigma$ = 0.48)\footnote{We have chosen to use the metrics gathered by the $P_f = 30\%$ filter since it is the best performing filter of those considered in this section that is able to find the injected faults for every experiment.}. However, if we only consider the larger projects where instrumentation is a more prevalent issue (\emph{i.e.} XML-Security and JMeter), the dynamic code coverage approach can reduce execution time by 27\% on average ($\sigma$ = 0.28). 

The other gathered metrics in this empirical evaluation, unlike execution time, show a consistent improvement over \ac{SFL} in every project. In average, the \ac{DCC} approach reduced 63\% ($\sigma$ = 0.30) the generated fault localization ranking, providing a more concise report when compared to \ac{SFL}. The quality of diagnosis, described in equation~\ref{eq:qd}, also suffered a slight improvement, from 85\% ($\sigma$ = 0.20) without \ac{DCC} to 87\% ($\sigma$ = 0.19) with \ac{DCC}.


\section{Threats to Validity} 
\label{sec:threats_to_validity}

The main threat to external validity of these empirical results is the fact that only four test subjects were considered. Although the subjects were all real, open source software projects, it is plausible to assume that a different set of subjects, having inherently different characteristics, may yield different results. Other threat to external validity is related to the injected faults used in the experiments. These injected faults, despite being fifteen in total for each experimental subject, may not represent the entire conceivable software fault spectrum. We are also assuming that the experimental subjects do not have any faults besides those that are injected by us, and that their test cases were correctly formulated and implemented.

Threats to internal validity are related to some fault in the \ac{DCC} implementation, or any underlying implementation, such as \ac{SFL} or even the instrumentation for gathering program spectra. To minimize this risk, some testing and individual result checking were performed before the experimental phase.


\acresetall
\chapter{Conclusions and Future Work} 
\label{cha:conclusions}

Currently, the most effective debugging tools are \acp{SDT}. During the
state of the art study, some flaws regarding efficiency were presented.
Afterwards, a solution was proposed to minimize these flaws -- the \ac{DCC}
approach and an empirical evaluation was performed to assess the validity of the
proposed solution.

\section{State of the art of Debugging Tools} 
\label{sec:state_of_the_art_of_debugging_tools}

After studying the currently available debugging methodologies, \acp{SDT} stand out as
the most effective tools to address fault localization. Other tools required more effort
by the user and a more in-depth knowledge of the \ac{SUT} in order to debug it. In fact,
with \acp{SDT}, minimal knowledge of the \ac{SUT} is required to locate faults.

\acp{SDT} also help to improve the software development process by automating fault 
localization and thus enabling regression testing.

Some inefficiencies were found in current \ac{SDT} implementations, mainly because of the
instrumentation overhead required to address fault localization. This would particularly 
affect large scale projects, because of their high quantity of \acp{LOC} that need to be
instrumented.

Other debugging techniques were also studied, namely model-based reasoning techniques, such as \ac{MBSD}. \ac{MBSD} can produce more accurate diagnostic hypotheses when compared with statistical approaches. However, the computational effort required to create a formal model of a large, real world software application remains highly prohibitive.


\section{Proposed Solution} 
\label{sec:proposed_solution}

The solution proposed in this thesis tries to avoid the \ac{LOC} level of instrumentation
detail in fault localization, while still using statistical debugging techniques.

This approach, named \ac{DCC}, would start by using a coarser granularity of instrumentation
and progressively increasing the instrumentation detail of certain components, based on
intermediate results provided by the \ac{SFL} technique.

In order to assess the validity of our approach, we have conducted an empirical evaluation on four real world open-source software projects. In each project, we have injected 15 different faults, and performed the fault localization task with \ac{SFL} and \ac{DCC}. With the empirical evaluation results we have demonstrated that, for large projects, our approach not only reduces the execution time by 27\% on average, but also reduces the number of components reported to the user by 63\% on average.

Let us revisit the research question presented in Section~\ref{sec:research_questions}:

\begin{itemize}
	\item {How can a fault localization approach that instruments less software components obtain similar diagnostic results when compared with \ac{SFL}, while reducing execution overhead?}
\end{itemize}

We can conclude that our approach meets the requirements of our research question. \ac{DCC} is a fault localization methodology that, by employing a dynamic, iterative approach, is able to instrument less software components when compared to other statistical fault localization techniques. Furthermore, the fault localization diagnostic reports generated by this methodology are often times shorter than \ac{SFL} and the execution time is also significantly reduced when \ac{DCC} is employed.


\section{Main Contributions} 
\label{sec:main_contributions}

This thesis makes the following main contributions:
\begin{enumerate}
\item \ac{DCC}, a technique that automatically decides the instrumentation granularity for each module in the system, has been proposed; and

\item An empirical study to validate the proposed technique, demonstrating its efficiency using real-world, large programs. The empirical results shows that \ac{DCC} can indeed decrease the overhead imposed in the software under test, while still maintaining the same diagnostic accuracy as current approaches to fault localization. \ac{DCC} also decreases the diagnostic report size when compared to traditional \ac{SFL}.
\end{enumerate}

To the best of our knowledge, our dynamic code coverage approach has not been described before.


\section{Publications} 
\label{sec:publications}

The \ac{DCC} algorithm and the GZoltar tool have also had some exposure in academia. With the work done in this thesis, we were able to submit to the following conferences: 

\begin{itemize}
	\item Alexandre Perez, André Riboira and Rui Abreu. \textbf{Fault Localization using Dynamic Code Coverage} -- submitted and accepted into \emph{The 5\textsuperscript{th} Meeting of Young Researchers of University of Porto (IJUP'12)}, 2012. This paper outlines the idea of using multiple levels of detail to mitigate instrumentation overhead in fault localization tools that use \ac{SFL}.
	\item José Campos, Alexandre Perez, André Riboira and Rui Abreu. \textbf{GZoltar: an Eclipse plug-in for Testing and Debugging} -- submitted and accepted into \emph{The 27\textsuperscript{th} IEEE/ACM International Conference on Automated Software Engineering (ASE'12) -- Tool Demonstration}, 2012. This tool demonstration aims to present the GZoltar toolset, as well as its underlying architecture.
\end{itemize}

As of this writing, another paper is being prepared for submission to \emph{The 6\textsuperscript{th} IEEE International Conference on Software Testing, Verification and Validation (ICST'13)}. This paper is authored by Alexandte Perez, André Riboira and Rui Abreu. The publication will describe the \ac{DCC} algorithm, and present an empirical evaluation of its performance compared to traditional \ac{SFL}. All of the publications mentioned above are compiled in Appendix~\ref{cha:publications}.


\section{Future Work} 
\label{sec:future_work}

Although the initial goals of this work were achieved, there are various subjects worth researching in order to further improve the \ac{DCC} fault localization methodology. Some aspects of the dynamic code coverage technique that still require further investigation are presented in the following paragraphs.

One subject worth investigating is the way of how the initial system granularity is established. Currently, this value is set manually and is the same across the entire system under test. A way to change this would be by using static analysis to assess program information and to adjust the system's initial granularity accordingly. Another approach would be to learn what were the most frequently expanded components from previous executions, and change these components' initial granularity independently. 

Other issue that requires further investigation pertains to the filtering methods. It is possible that there are better filtering methods than the ones presented in this paper, namely methods that employ dynamic strategies, that change the cutting threshold based on program spectra analysis.

Our empirical evaluation results were gathered by injecting single faults. While good results have been observed, they may not represent the real world software development environment, where multiple faults may arise. Further investigation is needed, then, to evaluate \ac{DCC} performance when tackling multiple simultaneous software faults. 

One may also explore the ability that \ac{DCC} has at stopping exploration at any desired granularity level. This methodology could be combined, then, with complementary fault localization methodologies, such as model-based reasoning approaches to debugging. As an example, \ac{DCC} can be employed to obtain the top ranked methods in a program, and then the \ac{MBSD} technique can be used only on these software components. \ac{DCC} would be acting as a pruning mechanism, restricting the exploration performed by \ac{MBSD}, and significantly reducing the amount of computation required by this model-based technique.  




\PrintBib{mieic_en}

\appendix

\chapter{Publications} 
\label{cha:publications}

\includepaper{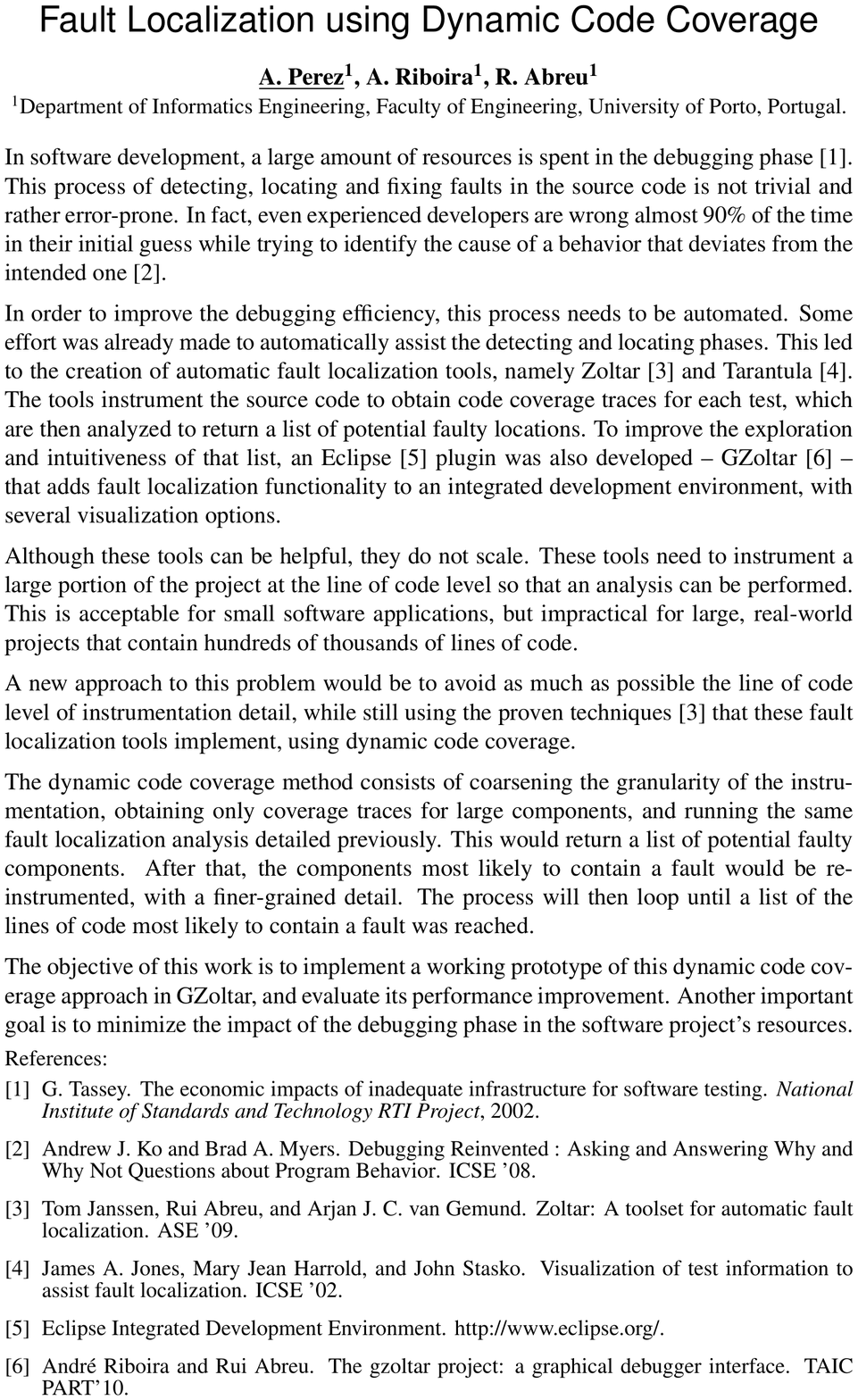}{Fault Localization using Dynamic Code Coverage}{ijup12}
\includepaper{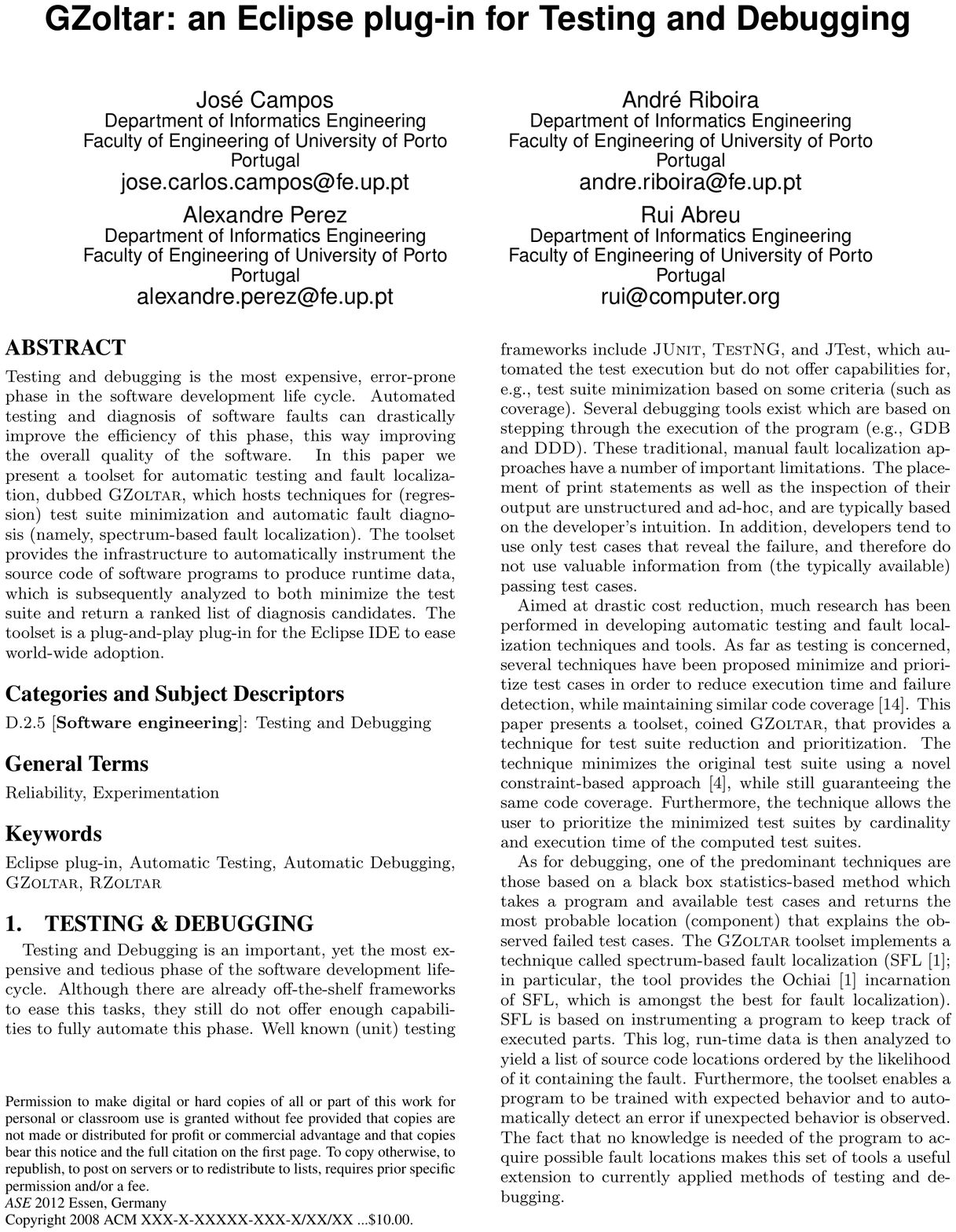}{GZoltar: an Eclipse plug-in for Testing and Debugging}{ase12}
\includepaper{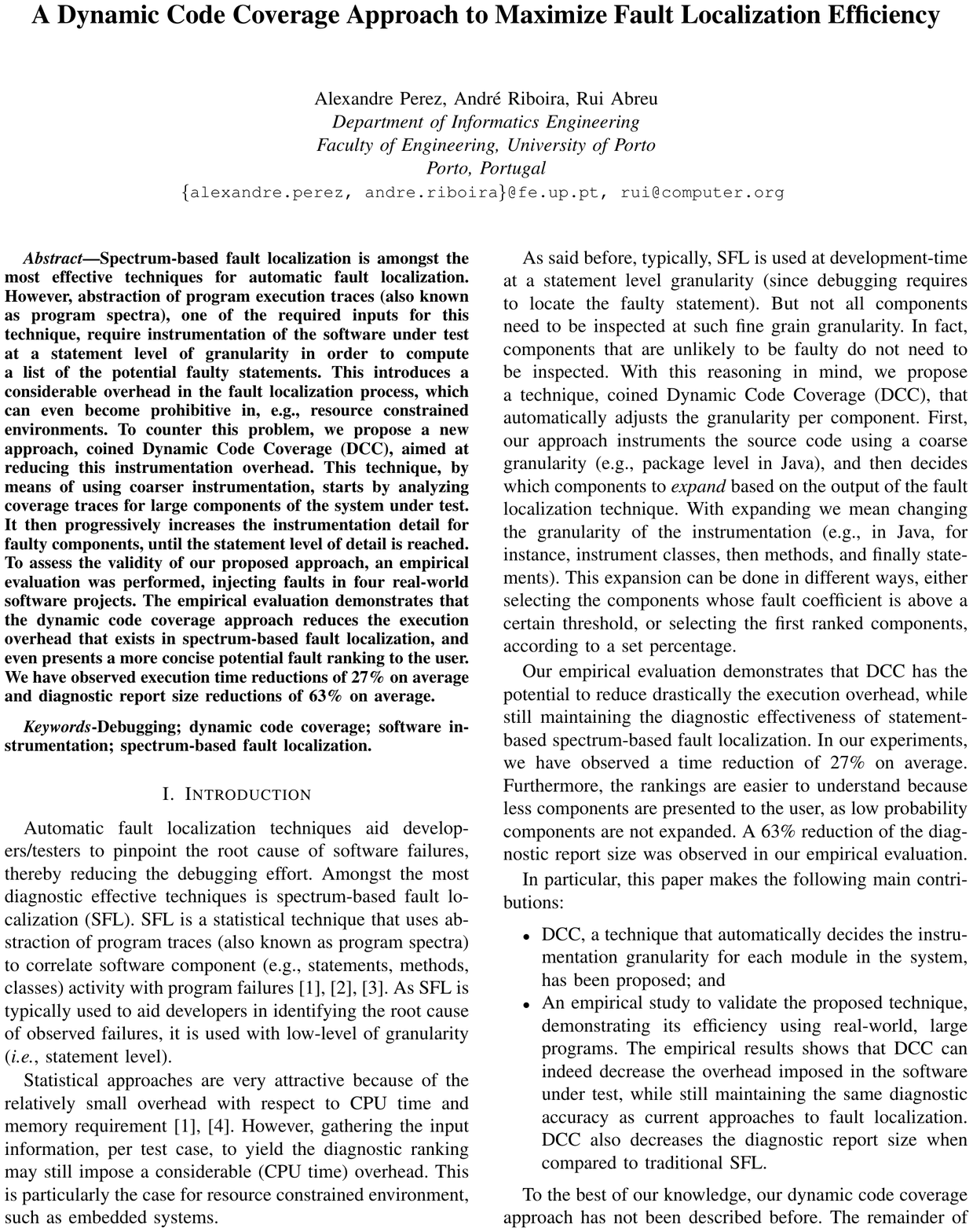}{A Dynamic Code Coverage Approach to Maximize Fault Localization Efficiency}{icst13}



\end{document}